\documentclass{article}
\usepackage{arxiv}
\usepackage{microtype}
\usepackage{graphicx}
\usepackage{booktabs} 
\usepackage{natbib}
\usepackage{float}
\usepackage[hyperfootnotes=false]{hyperref}       
\usepackage{amsmath}
\usepackage{amssymb}
\usepackage{mathtools}
\usepackage{amsthm}
\usepackage[capitalize,noabbrev]{cleveref}
\usepackage{graphicx, subcaption}
\usepackage[normalem]{ulem}
\usepackage[utf8]{inputenc} 
\usepackage[T1]{fontenc}    
\usepackage{url}            
\usepackage{booktabs}       
\usepackage{amsfonts}       
\usepackage{nicefrac}       
\usepackage{microtype}      
\usepackage{xcolor}         
\usepackage{bm}
\usepackage{pifont}
\usepackage[utf8]{inputenc}
\newcommand{\xmark}{\ding{55}}%
\usepackage{hhline}
\usepackage{multirow}
\usepackage{footmisc}

\newcommand{\fwd}[1]{\mathcal{A}(#1)}

\newcommand{\gaussian}[2][\vct{0}]{\mathcal{N}(#1, #2 \textbf{I})}

\newcommand{\norm}[1]{\left\lVert#1\right\rVert}

\newtheorem{definition}{Definition}
\newcommand{\vct}[1]{\bm{#1}}
\newcommand{\mtx}[1]{\bm{#1}}
\newcommand{\diff}{\text{d}}

\newcommand{\acumprod}{\bar{\alpha}_i}
\newcommand{\enc}[1]{\mathcal{E}_0(#1)}
\newcommand{\senc}[1]{\hat{\mathcal{E}}_{\vct{\theta}}(#1)}
\newcommand{\sencz}[1]{\hat{\mathcal{E}}_{\vct{z}}(#1)}
\newcommand{\sencs}[1]{\hat{\mathcal{E}}_{\sigma}(#1)}
\newcommand{\dec}[1]{\mathcal{D}_0(#1)}

\newcommand{\methodname}{Flash-Diffusion}
\newcommand{\methodnameexplain}{\uline{F}ast \uline{L}atent Sample-\uline{A}daptive Reconstruction \uline{S}c\uline{H}eme}
\newcommand{\adaptive}[1]{\texttt{Flash}(#1)}

\newcommand\blfootnote[1]{%
	\begingroup
	\renewcommand\thefootnote{}\footnote{#1}%
	\addtocounter{footnote}{-1}%
	\endgroup
}

\title{Adapt and Diffuse: Sample-adaptive Reconstruction via Latent Diffusion Models}

\author{Zalan Fabian\footnotemark[1] \quad Berk Tinaz\thanks{equal contribution} \quad Mahdi Soltanolkotabi \\
	University of Southern California\\
	Department of Electrical and Computer Engineering\\
	{\tt\small \{zfabian,tinaz,soltanol\}@usc.edu}
}
\begin{document}
\maketitle
\setcounter{footnote}{0} 
\begin{abstract}
Inverse problems arise in a multitude of applications, where the goal is to recover a clean signal from noisy and possibly (non)linear observations. The difficulty of a reconstruction problem  depends on multiple factors, such as the ground truth signal structure, the severity of the degradation and the complex interactions between the above. This results in natural sample-by-sample variation in the difficulty of a reconstruction problem. Our key observation is that most existing inverse problem solvers lack the ability to adapt their compute power to the difficulty of the reconstruction task, resulting in subpar performance and wasteful resource allocation. We propose a novel method, \textit{severity encoding},  to estimate the degradation severity of corrupted signals in the latent space of an autoencoder. We show that the estimated severity has strong correlation with the true corruption level and can provide useful hints on the difficulty of reconstruction problems on a sample-by-sample basis. Furthermore, we propose a reconstruction method based on latent diffusion models that leverages the predicted degradation severities to fine-tune the reverse diffusion sampling trajectory and thus achieve sample-adaptive inference times. Our framework, \methodname{}, acts as a wrapper that can be combined with any latent diffusion-based baseline solver, imbuing it with sample-adaptivity and acceleration. We perform experiments on both linear and nonlinear inverse problems and demonstrate that our technique greatly improves the performance of the baseline solver and achieves up to $10\times$ acceleration in mean sampling speed\footnote{Code is available at \url{https://github.com/z-fabian/flash-diffusion}}. 
\end{abstract}
\blfootnote{\textit{Published at the 41st International Conference on Machine Learning, Vienna, Austria, 2024}}
\section{Introduction}
Inverse problems arise in a multitude of computer vision \citep{ledig2017photo,wang2018image}, biomedical imaging  \citep{ sriram_end--end_2020} and scientific \citep{hand2018phase} applications, where the goal is to recover a clean signal from noisy and degraded observations. As information is fundamentally lost in the process, structural assumptions on the clean signal are needed to make recovery possible. Traditional compressed sensing \citep{candes_stable_2006} approaches utilize explicit regularizers that encourage sparsity of the reconstruction in transformation domains. More recently, data-driven deep learning methods have established new state-of-the-art in tackling most inverse problems (see an overview in \citep{ongie_deep_2020}).

A key shortcoming of available techniques is their inherent inability to adapt their compute power allocation to the difficulty of reconstructing a given corrupted sample. There is a natural sample-by-sample variation in the difficulty of recovery due to multiple factors. First, variations in the measurement process (e.~g.~more or less additive noise, different blur kernels) greatly impact the difficulty of reconstruction. Second, a sample can be inherently difficult to reconstruct for the particular model, if it is different from examples seen in the training set (out-of-distribution samples). Third, the amount of information loss due to the interaction between the specific sample and the applied degradation can vary vastly. For instance, applying a blur kernel to an image consisting of high-frequency textures destroys significantly more information than applying the same kernel to a smooth image. Finally, the implicit bias of the model architecture towards certain signal classes (e.g. piece-wise constant or smooth for convolutional architectures) can be a key factor in determining the difficulty of a recovery task. Therefore, expending the same amount of compute to reconstruct all examples is potentially wasteful, especially on datasets with varied corruption parameters.
 
  \begin{figure*}[t]
 	\centering
 	\includegraphics[width=0.75\textwidth]{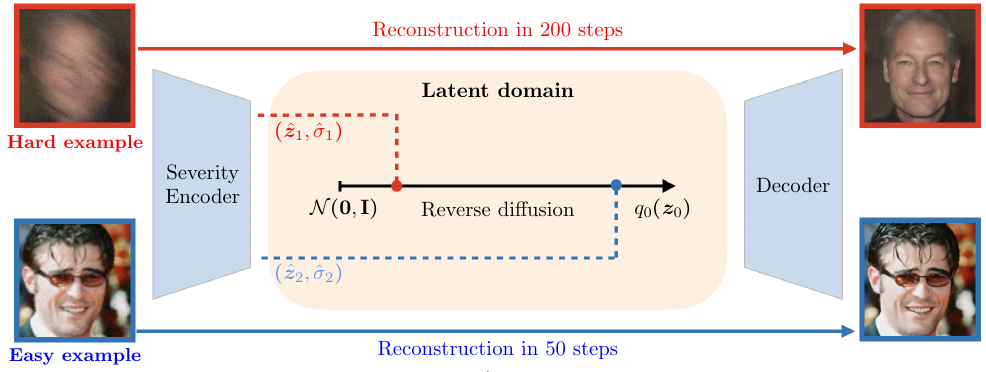}
 	\caption{\textbf{Overview of our method}: we estimate the degradation severity of corrupted images in the latent space of an autoencoder (Severity Encoder). We leverage the severity predictions ($\hat{\sigma}$) to find the optimal start time in a latent reverse diffusion process on a sample-by-sample basis. As a result, inference cost is automatically scaled by the difficulty of the reconstruction task at test time.}
 	\label{fig:front_page}
 \end{figure*}
 
 Sample-adaptive methods that incorporate the difficulty of a reconstruction problem, or \textit{the severity of degradation},  and allocate compute effort accordingly are thus highly desired. To the best of our knowledge, such methods have not been studied extensively in the literature. \textit{Unrolled networks} \citep{zhang_ista-net_2018, sun2016deep} have been proposed for reconstruction, that map the iterations of popular optimizers to learnable submodules, where deeper networks can be used to tackle more challenging reconstruction tasks. However, network size is determined in training time and therefore these methods are unable to adapt on a sample-by-sample basis. Deep Equilibrium Models have been proposed to solve inverse problems \citep{gilton2021deep} by training networks of arbitrary depth through the construction of fixed-point iterations. These methods can adapt their compute in test time by scaling the number of iterations to convergence, however it is unclear how the optimal number of iterations correlates with degradation severity.
 
 Diffusion models have established new state-of-the-art performance in synthesizing data of various modalities \citep{dhariwal_diffusion_2021, nichol2021glide, ramesh2022hierarchical, rombach2022high,  saharia2022photorealistic, ho2022cascaded, saharia2022palette, ho2022video, kong2020diffwave}, inverse problem solving and image restoration  \citep{kadkhodaie2021stochastic, saharia_image_2021, song2021solving, chung2022score, chung2022come, chung2022diffusion, chung2022improving,kawar2021snips, kawar2022denoising, kawar2022jpeg, fabian2023diracdiffusion}. Diffusion-based sampling techniques generate the missing information destroyed by the corruption step-by-step through a diffusion process that transforms pure noise into a target distribution. Recent work \citep{chung2022come} has shown that the sampling trajectory can be significantly shortened by starting the reverse diffusion process from a good initial reconstruction, instead of pure noise. However, this approach treats the noise level of the starting manifold as a hyperparameter independent of degradation severity. Therefore, even though sampling is accelerated, the same number of function evaluations are required to reconstruct any sample.
 
 
 More recently, latent domain diffusion, that is a diffusion process defined in the low-dimensional latent space of a pre-trained autoencoder, has demonstrated great success in image synthesis \citep{rombach2022high} and has been successfully applied to solving inverse problems \citep{rout2023solving, song2023solving, chung2023prompt} and in high-resolution \citep{luo2023refusion} and real-world \citep{jiang2023autodir} image restoration. Latent diffusion has the clear benefit of improved efficiency due to the reduced dimensionality of the problem leading to faster sampling. In addition to this, the latent space consists of compressed representations of relevant information in data and thus provides a natural space to quantify the loss of information due to image corruptions, which strongly correlates with the difficulty of the reconstruction task.
 
 In this paper, we propose a novel reconstruction framework (Figure \ref{fig:front_page}), where the cost of inference is automatically scaled based on the difficulty of the reconstruction task on a sample-by-sample basis. Our contributions are as follows:
 \begin{itemize}
 	\item We propose a novel method that we call \textit{severity encoding},  to estimate the degradation severity of noisy, degraded images in the latent space of an autoencoder. We show that the estimated severity has strong correlation with the true corruption level and can give useful hints at the difficulty of reconstruction problems on a sample-by-sample basis. Training the severity encoder is efficient, as it can be done by fine-tuning a pre-trained encoder.
 	\item We propose a technique for solving general noisy inverse problems based on latent diffusion models that leverages the predicted degradation severities to fine-tune the reverse diffusion sampling trajectory and thus achieve sample-adaptive inference times. Our framework is extremely flexible: it acts as a wrapper that can be combined with any latent diffusion-based inverse problem solver, imbuing it with sample-adaptivity and acceleration. We call our method \methodname{}: \methodnameexplain.
 	\item We perform in-depth numerical experiments on multiple datasets and on both linear and nonlinear noisy inverse problems. We demonstrate that (1) \methodname{} greatly improves the reconstruction performance of the baseline solver, especially in case of degradations with highly varying severity, and (2) \methodname{} accelerates the baseline solver by up to $10 \times$ without any loss in reconstruction quality.
 \end{itemize}
\section{Background}
\textbf{Diffusion models --} Diffusion in the context of generative modeling consists of transforming a clean data distribution $\vct{x}_0 \sim q_0(\vct{x})$ through a  forward noising process, defined over $0 \leq t \leq T, ~ t\in \mathbb{R}$, into some tractable distribution $q_T$. Typically, $q_t$ is chosen such that $\vct{x}_t$ is obtained from $\vct{x}_0$ via adding \textit{i.i.d.} Gaussian noise, that is $q_t(\vct{x}_t | \vct{x}_0) \sim \mathcal{N}(\vct{x}_0, \sigma_t^2 \mathbf{I})$, where $\sigma_t^2$ is from a known variance schedule. Diffusion models (DMs) \citep{sohl2015deep, ho_denoising_2020, song_generative_2020, song_improved_2020} learn to reverse the forward corruption process in order to generate data samples starting from a simple Gaussian distribution. The forward process can be described as an Itô stochastic differential equation (SDE) \citep{song2020score} $	\diff \vct{x} = f(\vct{x}, t) \diff t + g(t) \diff \vct{w},$
where $\vct{w} \in \mathbb{R}^n$ is the standard Wiener process. The forward SDE is reversible \citep{anderson1982reverse} and can be simulated given $\nabla_{\vct{x}} \log q_t(\vct{x})$, which is referred to as the \textit{score} of the data distribution. The score is typically approximated by a neural network $s_{\vct{\theta}}(\vct{x}_t, t)$ and learned from data. After discretizing the reverse SDE, a variety of discrete time sampling algorithms can be derived.

 Denoising Diffusion Probabilistic Models (DDPMs) \citep{sohl2015deep, ho_denoising_2020} are obtained from the discretization of the variance preserving SDE with $ f(\vct{x}, t) = -\frac{1}{2} \beta_t  \vct{x}$ and $g(t) = \sqrt{\beta_t}$, where $\beta_t$ is a pre-defined variance schedule that is a strictly increasing function of $t$. One can sample from the corresponding forward diffusion process at any time step $i$ as $\vct{x}_i =\sqrt{\acumprod} \vct{x}_0 + \sqrt{1 - \acumprod}\vct{\varepsilon},$
with $\vct{\varepsilon} \sim \mathcal{N}(\mathbf{0}, \mathbf{I})$ and $\acumprod = \prod_{j=1}^{i} \alpha_i, ~ \alpha_i = 1 - \beta_i$. By minimizing the denoising score-matching objective 
\begin{equation*}
	L_{DM} = \mathbb{E}_{\varepsilon, i, \vct{x}_i \sim q_0(\vct{x}_0) q_i(\vct{x}_i|\vct{x}_0)} \left[\norm{\varepsilon_{\vct{\theta}}(\vct{x}_i, i) - \varepsilon}^2 \right]
\end{equation*}
the score model $\varepsilon_{\vct{\theta}}(\vct{x}_i, i)$ learns to predict the noise on the input corrupted signal.





\textbf{Latent Diffusion Models (LDMs) --} LDMs \citep{rombach2022high} aim to mitigate the computational burden of traditional DMs by running diffusion in a low-dimensional latent space of an autoencoder. In particular, an encoder $\mathcal{E}_0$  is trained to extract a compressed representation $\vct{z} \in \mathbb{R}^d,~ d << n$ of the input signal $\vct{x}$ in the form $\vct{z} = \enc{\vct{x}}$. To recover the clean signal from the latent representation $\vct{z}$, a decoder $\mathcal{D}_0$ is trained such that $\dec{\enc{\vct{x}}} \approx \vct{x}$.  A score model that progressively denoises $\vct{z}$ can be trained in the latent space of the autoencoder 
via the objective 
following the DDPM framework. The final generated image can be obtained by passing the denoised latent through $\mathcal{D}_0$.



\textbf{Diffusion models for solving inverse problems --} Solving a general noisy inverse problem amounts to finding the clean signal $\vct{x} \in \mathbb{R}^n$ from a noisy and degraded observation $\vct{y} \in \mathbb{R}^m$ in the form
\begin{equation}\label{eq:inverse_problem}
	\vct{y} = \fwd{\vct{x}} + \vct{n},
\end{equation}
where $\mathcal{A}:\mathbb{R}^n\rightarrow\mathbb{R}^m$ denotes a deterministic degradation and $\vct{n} \sim \gaussian{\sigma_y^2}$ is Gaussian noise. As information is fundamentally lost in the measurement process, structural assumptions on clean signals are necessary to recover $\vct{x}$. Deep learning approaches can learn a generative model $p_\theta(\vct{x})$ that represents the underlying structure of clean data and can be leveraged as a prior to solve  \eqref{eq:inverse_problem}. In particular, the posterior over clean data can be written as $p_\theta(\vct{x} | \vct{y}) \propto p_\theta(\vct{x}) p(\vct{y} | \vct{x})$, where the likelihood $ p(\vct{y} | \vct{x})$ is represented by \eqref{eq:inverse_problem}. Thus, one can sample from the posterior by querying the generative model. The score of the posterior can be written as 
\begin{equation*}
	\nabla_{\vct{x}}  \log p_\theta(\vct{x} | \vct{y}) = \nabla_{\vct{x}}  \log p_\theta(\vct{x}) + \nabla_{\vct{x}}  \log p(\vct{y} | \vct{x}),
\end{equation*}
where the first term corresponds to an unconditional score model trained to predict noise on the signal without any information about the forward model $\mathcal{A}$. The score of the likelihood term however is challenging to estimate in general. Various approaches have emerged to incorporate the data acquisition model into a diffusion process, including projection-based approaches \citep{song2021solving, chung2022score, chung2022come}, restricting updates to stay on a given manifold \citep{chung2022improving}, spectral approaches \citep{kawar2022denoising}, or methods that tailor the diffusion process to the degradation \citep{welker2022driftrec, fabian2023diracdiffusion, delbracio2023inversion}. \citet{chung2022diffusion} proposes Diffusion Posterior Sampling (DPS) that approximates the gradient of the likelihood as $\nabla_{\vct{x}_i}  \log p(\vct{y} | \vct{x}_i) \approx \nabla_{\vct{x}_i}  \log p(\vct{y} | \hat{\vct{x}}_0(\vct{x}_i))$, where $\hat{\vct{x}}_0(\vct{x}_i)$ is the posterior mean estimate obtained from the score model.

More recently, a flurry of activity has emerged to deploy LDMs for inverse problem solving. The most straightforward approach is to perform posterior sampling in latent space via the DPS approximation 
\begin{equation*}
\nabla_{\vct{z}_i}  \log p(\vct{y} | \vct{z}_i) \approx \nabla_{\vct{z}_i}  \log p(\vct{y} | \mathcal{D}_0(\hat{\vct{z}}_0(\vct{z}_i))),
\end{equation*}
 which can be viewed as a "vanilla" extension of DPS to the latent domain  \citep{rout2023solving} . We referred to this technique as Latent-DPS following \citet{song2023solving}. However, Latent-DPS achieves poor performance out-of-the-box due to the inaccuracy of the approximation in early stages of diffusion and due to additional complications introduced by the decoder \citep{rout2023solving, song2023solving, chung2023prompt}. To address the latter concern, \citet{rout2023solving} proposes two extensions of Latent-DPS: GML-DPS, that guides the diffusion towards fixed points of the encoder-decoder composition, and PSLD which adds a "gluing" term in order to avoid inconsistencies at mask boundaries. Authors in \citet{song2023solving} take a different route and propose ReSample, a sampling technique that enforces data consistency directly on the posterior means by solving the optimization problem 
$$ \hat{\vct{z}}_0(\vct{y}) = \text{arg}\min_{\vct{z}} \| \vct{y} - \fwd{\mathcal{D}_0(\vct{z})}\|_2^2 + \lambda \|\vct{z} -\hat{\vct{z}}_0(\vct{z}_i) \|_2^2,$$ or its pixel domain equivalent, both referred to as hard data consistency. \citet{chung2023prompt} leverages text-to-image LDMs and incorporates prompt-tuning into their algorithm while combining Latent-DPS and hard data consistency.

A key challenge of diffusion-based solvers is their heavy compute demand, as reconstructing a single sample requires typically $500-1000$ evaluations of a large score model. Come-Closer-Diffuse-Faster (CCDF) \citep{chung2022come} shortens the sampling trajectory by leveraging a good initial posterior mean estimate $\hat{\vct{x}}_0$ from a reconstruction network. They initialize the reverse process by jumping to a fixed time step in the forward process via $	\vct{x}_{k} =\sqrt{\bar{\alpha}_k} \hat{\vct{x}}_0 + \sqrt{1 - \bar{\alpha}_k}\vct{\varepsilon},$
and only perform $k << N$ reverse diffusion steps, where $k$ is a fixed hyperparameter. More recently, \citet{rout2023beyond} proposes leveraging an efficient second-order approximation for estimating the posterior mean, notably accelerating sampling.

\section{Method}
\subsection{Severity encoding}
The goal of inverse problem solving is to recover the clean signal $\vct{x}$ from a corrupted observation $\vct{y}$ (see \eqref{eq:inverse_problem}).  The degradation $\mathcal{A}$ and additive noise $\vct{n}$ fundamentally destroy information in $\vct{x}$. The amount of information loss, or the \textit{severity} of the degradation, strongly depends on the interaction between the signal structure and the specific degradation. For instance, blurring removes high-frequency information, which implies that applying a blur kernel to an image with abundant high-frequency detail (textures, hair, background clutter etc.) results in a \textit{more severe degradation} compared to applying the same kernel to a smooth image with few details. In other words, the difficulty of recovering the clean signal does not solely depend on the degradation process itself, but also on the specific signal the degradation is applied to. Thus, tuning the reconstruction method's capacity purely based on the forward model misses a key component of the problem: the data itself. 


Quantifying the severity of a degradation is a challenging task in image domain. As an example, consider the forward model $\vct{y} = c \vct{x}, ~c \in \mathbb{R}^+$ that simply rescales the clean signal. Recovery of $\vct{x}$ from $\vct{y}$ is trivial, however image similarity metrics such as PSNR or NMSE that are based on the Euclidean distance in image domain may indicate arbitrarily large discrepancy between the degraded and clean signals. On the other hand, consider $\vct{y} = \vct{x} + \vct{n}, ~ \vct{n} \sim \gaussian{\sigma^2}$ where the clean signal is simply perturbed by some additive random noise. Even though the image domain perturbation is (potentially) small, information is fundamentally lost and perfect reconstruction is no longer possible.

What is often referred to as the \textit{manifold hypothesis} \citep{bengio2013representation} states that natural images live in a lower dimensional manifold embedded in $n$-dimensional pixel-space. This in turn implies that the information contained in an image can be represented by a low-dimensional latent vector that encapsulates the relevant features of the image. Autoencoders \citep{kingma2013auto, razavi2019generating} learn a latent representation from data by first summarizing the input image into a compressed latent vector $\vct{z} =\enc{\vct{x}}$ through an encoder. Then, the original image can be recovered from the latent via the decoder $\hat{\vct{x}} = \dec{\vct{z}}$ such that $\vct{x} \approx \hat{\vct{x}}$. As the latent space of autoencoders contains only the relevant information of data, it is a more natural space to quantify the loss of information due to the degradation than the image domain.

In particular, assume that we have access to the latent representation of clean images $\vct{z}_0 = \enc{\vct{x}_0},~ \vct{z}_0 \in \mathbb{R}^d$, for instance from a pre-trained autoencoder. We propose a \textit{severity encoder} $\hat{\mathcal{E}}_{\vct{\theta}}$ that achieves two objectives simultaneously: (1) it can predict the latent representation of a clean image, given a noisy and degraded observation and (2) it can quantify the error in its own latent estimation. We denote 
$$\senc{\vct{y}} = (\hat{\vct{z}}, ~ \hat{\sigma}):= (\sencz{\vct{y}}, \sencs{\vct{y}} ),$$ 
with $\hat{\vct{z}} \in \mathbb{R}^d$ the estimate of $\vct{z}_0$ and $\hat{\sigma} \in \mathbb{R}$ the \textit{estimated degradation severity} to be specified shortly. We use the notation $\sencz{\vct{y}} = \hat{\vct{z}}$ and $\sencs{\vct{y}} = \hat{\sigma}$ for the two conceptual components of our model, however in practice a single architecture is used to represent $\hat{\mathcal{E}}_{\vct{\theta}}$. The first objective can be interpreted as image reconstruction in the latent space of the autoencoder: for $\vct{y} = \fwd{\vct{x}} + \vct{n}$ and $\vct{z}_0 = \enc{\vct{x}}$, we have $\sencz{\vct{y}} = \hat{\vct{z}} \approx \vct{z}_0$. The second objective captures the intuition that recovering $\vct{z}_0$ from $\vct{y}$ exactly may not be possible, and the prediction error is proportional to the loss of information about $\vct{x}$ due to the corruption. Thus, even though the predicted latent $\hat{\vct{z}}$ might be away from the true $\vct{z}_0$, the encoder quantifies the uncertainty in its own prediction. More specifically, we make the assumption that the prediction error in latent space can be modeled as zero-mean \textit{i.i.d.} Gaussian. That is, we assume that 
$$ \hat{\vct{z}}(\vct{y}) - \vct{z}_0 := \vct{e}(\vct{y}) \sim \mathcal{N}(\mathbf{0}, \sigma_*^2(\vct{y}) \mathbf{I}),$$
and we interpret the variance in prediction error $\sigma_*^2$ as the measure of degradation severity. We optimize the joint objective
\begin{equation}\label{eq:sev_loss}
 \mathbb{E}_{\vct{x}_0 \sim q_0(\vct{x}_0), \vct{y} \sim \mathcal{N}(\fwd{\vct{x}_0}, \sigma_y^2 \mathbf{I})} \bigg[ \norm{\vct{z}_0 - \sencz{\vct{y}}}^2 + 
 \lambda_{\sigma} \norm{\bar{\sigma}^2(\vct{y}, \vct{z}_0) - \sencs{\vct{y}}}^2 \bigg] := L_{lat. rec.} + \lambda_\sigma L_{err.},
\end{equation}
with $\vct{z}_0 = \enc{\vct{x}_0}$ for a fixed, pre-trained encoder $\mathcal{E}_0$ and 
\begin{equation*}
	\bar{\sigma}^2(\vct{y}, \vct{z}_0) = \frac{1}{d -1} \sum_{i = 1}^d (\vct{e}^{(i)} - \frac{1}{d} \sum_{j=1}^{d}\vct{e}^{(j)} )^2
\end{equation*}
is the sample variance of the prediction error estimating $\sigma_*^2$. Here $\lambda_{\sigma}>0$ is a hyperparameter that balances between reconstruction accuracy ($ L_{lat. rec.}$) and error prediction performance ($L_{err.}$).  We empirically observe that even small loss values of $L_{lat. rec.}$ (that is fairly good latent reconstruction) may correspond to visible reconstruction error in image domain as semantically less meaningful features in image domain are often not captured in the latent representation. Furthermore, small errors in latent space can be severely amplified in image space after decoding, depending on the smoothness of the pretrained decoder. Therefore, we utilize an extra loss term that imposes image domain consistency with the ground truth image in the form
\begin{equation*}
	L_{im. rec.} = \mathbb{E}_{\vct{x}_0 , \vct{y}} \left[ \norm{ \vct{x}_0 - \dec{\sencz{\vct{y}}}}^2\right],
\end{equation*}
resulting in the final combined loss 
$
		L_{sev}  =  L_{lat. rec.} + \lambda_\sigma L_{err.} + \lambda_{im.} L_{im. rec.} ,
$
with $ \lambda_{im.} \geq 0$ hyperparameter. Training the severity encoder is fast, as one can fine-tune the pre-trained encoder $\mathcal{E}_0$. 

\subsection{Sample-adaptive inference} 
\begin{figure}
	\centering
\begin{minipage}{.45\textwidth}
	\centering
	\includegraphics[width=0.89\linewidth]{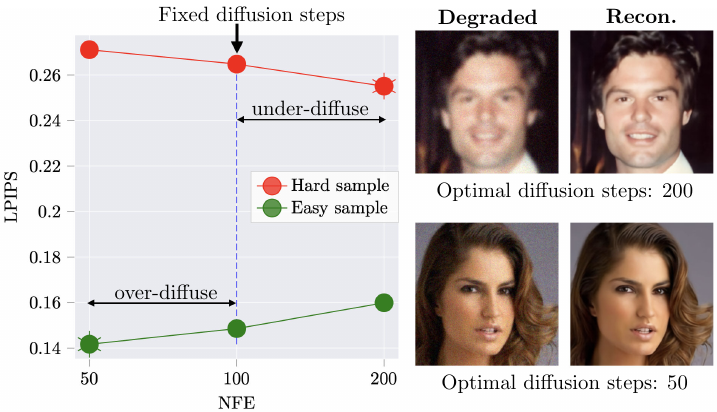}
	\caption{The optimal number of reverse diffusion steps varies depending on the severity of degradations. Fixing the number of steps results in over-diffusing some samples, whereas others could benefit from more iterations. 	\label{fig:optimal_t_diff}}
\end{minipage}\hspace{0.2cm}
\begin{minipage}{.45\textwidth}
	\centering
	\includegraphics[width=0.99\linewidth]{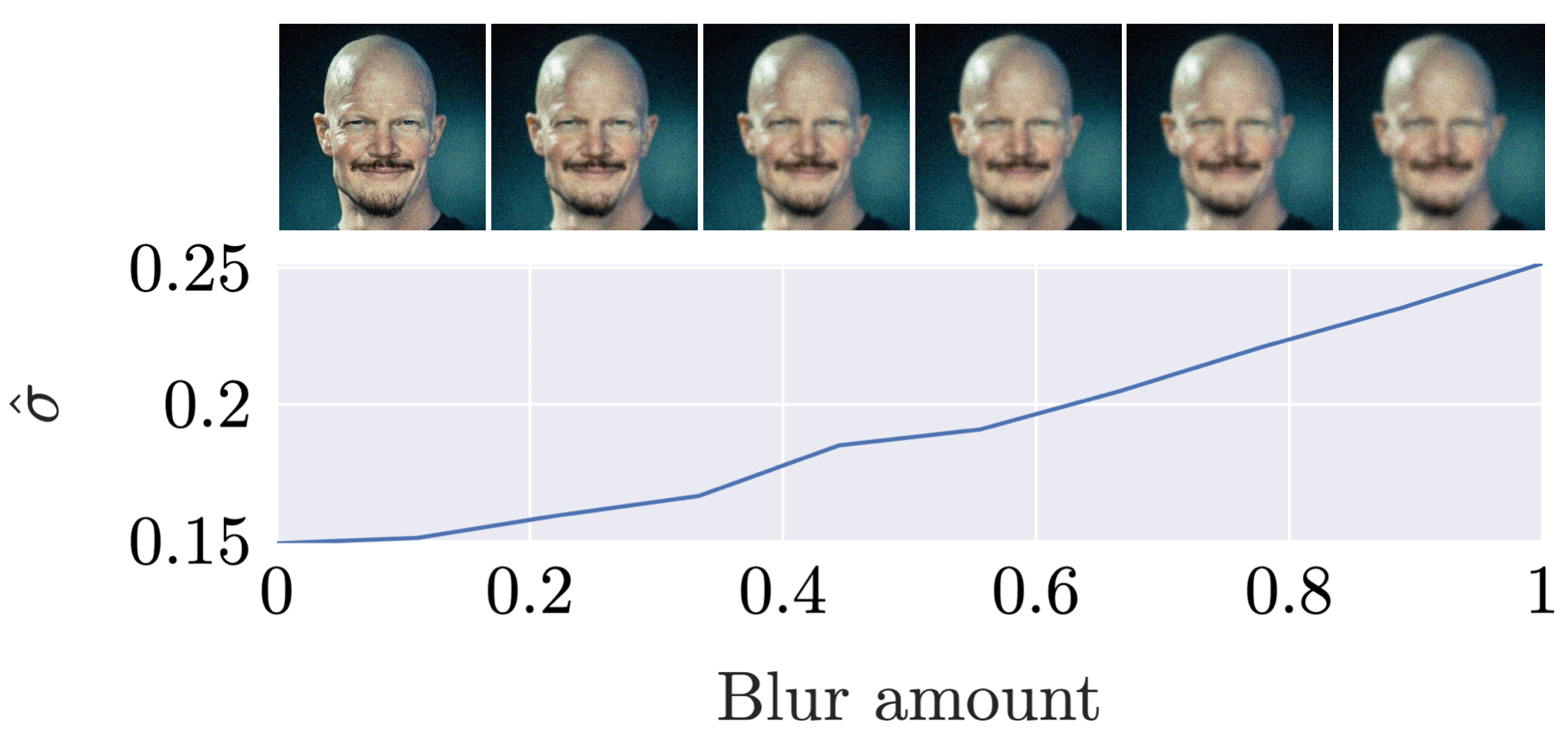}
	\caption{Effect of degradation on predicted severities: given a ground truth image corrupted by varying amount of blur, $\hat{\sigma}$ is a non-decreasing function of the blur amount.\label{fig:ordering_plot}}
\end{minipage}
\end{figure}

Diffusion-based inverse problem solvers synthesize missing data that has been destroyed by the degradation process through diffusion.  As shown in Figure \ref{fig:optimal_t_diff}, depending on the amount of missing information (easy vs. hard samples), the optimal number of diffusion steps may greatly vary. Too few steps may not allow the diffusion process to generate realistic details on the image, leaving the reconstruction overly smooth. On the other hand, diffusion-based solvers are known to hallucinate details that may be inconsistent with the ground truth signal, or even become unstable when too many reverse diffusion steps are applied. Authors in \citet{chung2022come} observe that there always exists an optimal spot between $0$ and $N$ diffusion steps that achieves the best reconstruction performance. We aim to automatically find this "sweet spot" on a sample-by-sample basis.

Our proposed severity encoder learns to map degraded signals to a noisy latent representation, where the noise level is proportional to the degradation severity. This provides us the unique opportunity to leverage a latent diffusion process to progressively denoise the latent estimate obtained from our encoder. Even more importantly, we can automatically scale the number of reverse diffusion steps required to reach the clean latent manifold based on the predicted severity.

\textbf{Finding the optimal starting time --} We find the time index $i_{start}$ in the latent diffusion process at which the signal-to-noise ratio (SNR) matches the SNR predicted by the severity encoder. Assume that the latent diffusion process is specified by the conditional distribution $q_i(\vct{z}_i | \vct{z}_0) \sim \mathcal{N}(a_i \vct{z}_0 , b_i^2 \mathbf{I})$, where $a_i$ and $b_i$ are determined by the specific sampling method (e. g. $a_i = \sqrt{\acumprod}$ and $b_i^2 = 1 - \acumprod$ for DDPM). On the other hand, we have the noisy latent estimate  $\hat{\vct{z}} \sim \mathcal{N}( \vct{z}_0 , \sigma_*^2(\vct{y}) \mathbf{I})$, where we estimate $\sigma_*^2$ by $\sencs{\vct{y}}$. Then, SNR matching gives us the starting time index
\begin{equation}\label{eq:snr_match}
	i_{start}(\vct{y}) = \arg \min_{i\in [1, 2, .., N]} \left| \frac{a_i^2}{b_i^2} - \frac{1}{\sencs{\vct{y}}} \right|.
\end{equation}
Thus, we start reverse diffusion from the initial reconstruction $\hat{\vct{z}}$ provided by the severity encoder and progressively denoise it using a pre-trained unconditional score model, where the length of the sampling trajectory is directly determined by the predicted severity of the degraded example. 

\textbf{Noise correction --} Even though we assume that the prediction error in latent space is \textit{i.i.d.} Gaussian in order to quantify the estimation error by a single scalar, in practice the error often has some structure. This can pose a challenge for the score model, as it has been trained to remove isotropic Gaussian noise. We observe that it is beneficial to mix $\hat{\vct{z}}$ with some \textit{i.i.d.} \textit{correction noise} in order to suppress structure in the prediction error. In particular, we initialize the reverse process by
\begin{equation*}
	\vct{z}_{start} =\sqrt{1 -  c \hat{\sigma}^2 }\hat{\vct{z}} + \sqrt{ c \hat{\sigma}^2}\vct{\varepsilon}, ~~ \vct{\varepsilon} \sim \mathcal{N}(\mathbf{0}, \mathbf{I}),
\end{equation*}
where $c \geq 0$ is a tuning parameter. We discuss alternatives to this initialization scheme based on adaptive DDIM encoding in Appendix \ref{apx:init}.

\section{Experiments}\label{sec:exp}
\begin{figure*}[t]
	\centering
	\includegraphics[width=0.7\linewidth]{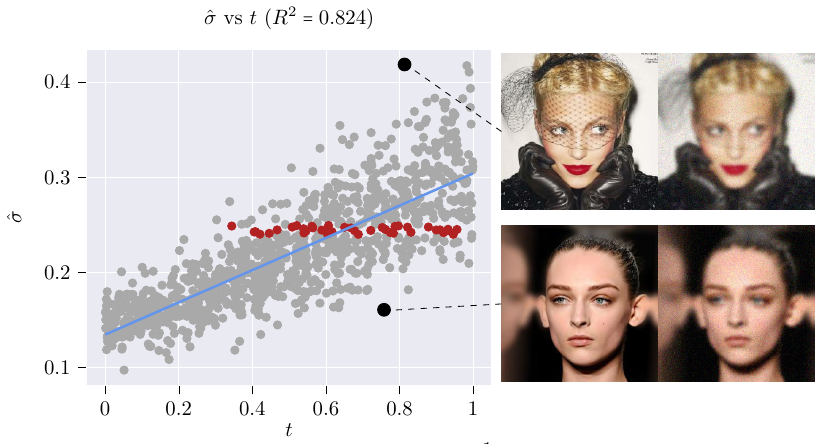}
	\label{fig:outliers}
	\caption{Blur amount ($t$) vs. predicted degradation severity ($\hat{\sigma}$).  Outliers indicate that the predicted degradation severity is not solely determined by the amount of blur. The bottom image is \textit{surprisingly easy} to reconstruct, as it is overwhelmingly smooth with features close to those seen in the training set. The top image is \textit{surprisingly hard}, due to more high-frequency details and unusual features not seen during training. Points in red suggest that a given degradation severity may result from a wide range of blur levels (see Fig. \ref{fig:contributors} and discussion under \textit{Identifying contributors to severity in Section \ref{sec:sev_enc}}).}
	\label{fig:sev_contrib}
\end{figure*}

\textbf{Dataset --} We perform experiments on CelebA-HQ ($256 \times 256$) \citep{karras_progressive_2018} and LSUN Bedrooms \citep{yu2015lsun}. We match the training and validation splits used in \citet{rombach2022high}, and set aside $200$ images from the validation split for testing. For comparisons involving the CelebA dataset and image domain score models we test on FFHQ \citep{karras2019style}, as pre-trained image-domain score models have been trained on the entire CelebA-HQ.

\textbf{Degradations --} We consider three degradations of diverging characteristics. \uline{Gaussian blur}: we apply Gaussian blur with kernel size of $61$. In order to investigate the effect of strongly varying degradation severity on sampling techniques, we analyze two settings: (1) varying Gaussian blur, where we sample the kernel standard deviation uniformly on $[0 , 3]$ ($0$ corresponds to no blurring) and (2) fixed Gaussian blur where the kernel standard deviation is set to $3$.  \uline{Nonlinear blur}: we deploy GOPRO motion blur simulated by a neural network model from \citet{tran2021explore}. This is a nonlinear forward model due to the camera response function. We randomly sample nonlinear blur kernels for each image. \uline{Varying random inpainting}: we randomly mask $70-80\%$ of the pixels, where the masking ratio is drawn uniformly. In all experiments, we add Gaussian noise to images in the $[0, 1]$ range with noise std of $0.05$.

\textbf{Baseline solvers and comparison methods --} \methodname{} acts as a wrapper around any latent diffusion solver adding sample-adaptivity and acceleration to the baseline inverse problem solver. We experiment with the following baseline solvers: Latent-DPS \citep{rout2023solving}, a straightforward adaptation of DPS \citep{chung2022diffusion} to the latent domain; GML-DPS and PSLD (linear-only) \citep{rout2023solving}, algorithms that are extensions of Latent-DPS with improved performance; and ReSample \citep{song2023solving}, a state-of-the-art latent diffusion solver leveraging hard data consistency. We compare the baseline solvers to their adaptive version obtained from our framework, which we denote by \adaptive{baseline solver}. Furthermore, we compare the results with DPS \citep{chung2022diffusion}, a well-established diffusion solver that operates in the pixel space. As our framework requires fine-tuning to the specific degradation, we additionally compare with SwinIR \citep{liang_swinir_2021}, a strong Transformer-based supervised image restoration model trained on corrupted-clean data pairs. Finally, we show results of decoding the severity encoder's $\hat{\vct{z}}$ estimate directly without diffusion, denoted by AE (autoencoded). Further details on comparison methods are in Appendix \ref{apx:experimental_details}.


\textbf{Models --} We use pre-trained DMs from \citet{dhariwal_diffusion_2021} for DPS and from \citet{rombach2022high} for LDMs. We fine-tune severity encoders from pre-trained LDM encoders and utilize a single convolution layer on top of $\hat{\vct{z}}$ to predict $\hat{\sigma}$. For more details on the experimental setup and hyperparameters, see Appendix \ref{apx:experimental_details}.

\textbf{Metrics --} We evaluate image distortion based on PSNR and SSIM and perceptual image quality based on LPIPS and FID. To evaluate speedup of adaptive solvers compared to their baseline, we compare the average number of reverse diffusion steps across the test dataset, denoted by $\overline{\text{NFE}}$.

\subsection{Severity encoding \label{sec:sev_enc}}
In this section, we investigate properties of the predicted degradation severity $\hat{\sigma}$. We perform experiments on a $1k$-image subset of the validation split. 

\textbf{Effect of degradation level --} First, we isolate the effect of degradation on $\hat{\sigma}$  (Fig. \ref{fig:ordering_plot}). We fix the clean image and apply increasing amount of Gaussian blur. We observe that $\hat{\sigma}$ is an increasing function of the blur amount applied to the image: heavier degradations on a given image result in higher predicted degradation severity. This implies that the severity encoder learns to capture the amount of information loss caused by the degradation.  

\textbf{Interaction between degradation and image --} Next, we investigate the relation between $\hat{\sigma}$ and the underlying degradation severity (Fig. \ref{fig:sev_contrib}). We parameterize the corruption level by $t$, where $t=0$ corresponds to no blur and $t=1$ corresponds to the highest blur level where in both cases we add Gaussian noise ($\sigma=0.05$). We vary the blur kernel width linearly for $t \in (0, 1)$. We observe that the predicted $\hat{\sigma}$ severities strongly correlate with the corruption level. However, the existence of outliers suggest that factors other than the corruption level may also contribute to the predicted severities. The bottom image is predicted to be \textit{surprisingly easy}, as other images of the same corruption level are typically assigned higher predicted severities. This sample is overwhelmingly smooth, with a lack of fine details and textures, such as hair, present in other images. Moreover, the image shares common features with others in the training set. On the other hand, the top image is considered \textit{surprisingly difficult}, as it contains unexpected features and high-frequency details that are uncommon in the dataset. This example highlights the potential application of our technique to hard example mining and dataset distillation. 

\begin{figure}[ht]
	\centering
	\begin{subfigure}{0.7\linewidth}
		\centering
		\includegraphics[width=1.0\linewidth]{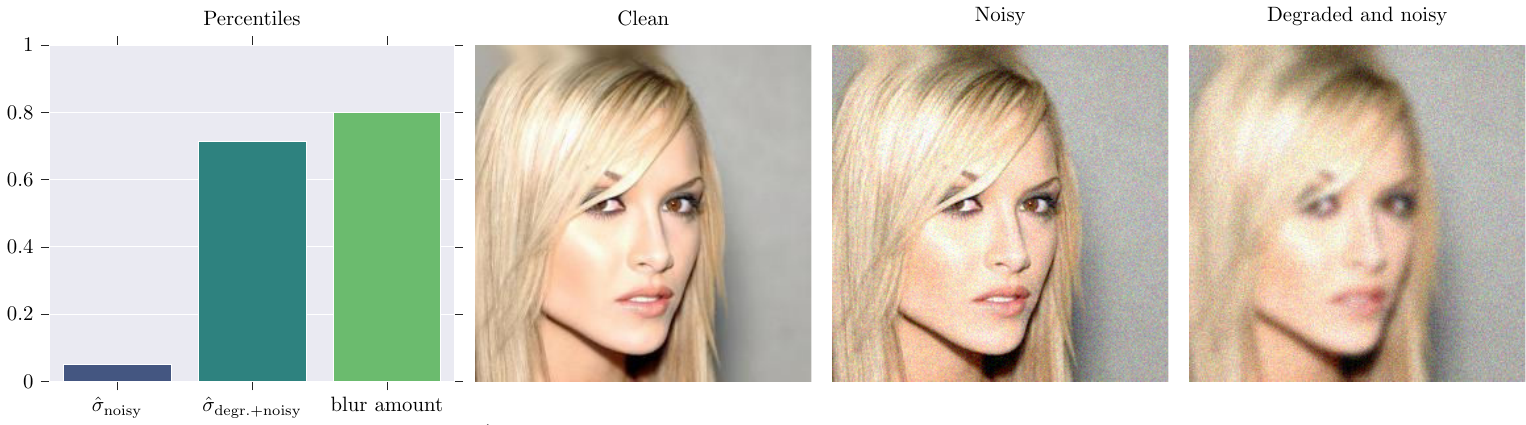}
	\end{subfigure}
	\begin{subfigure}{0.7\linewidth}
		\centering
		\includegraphics[width=1.0\linewidth]{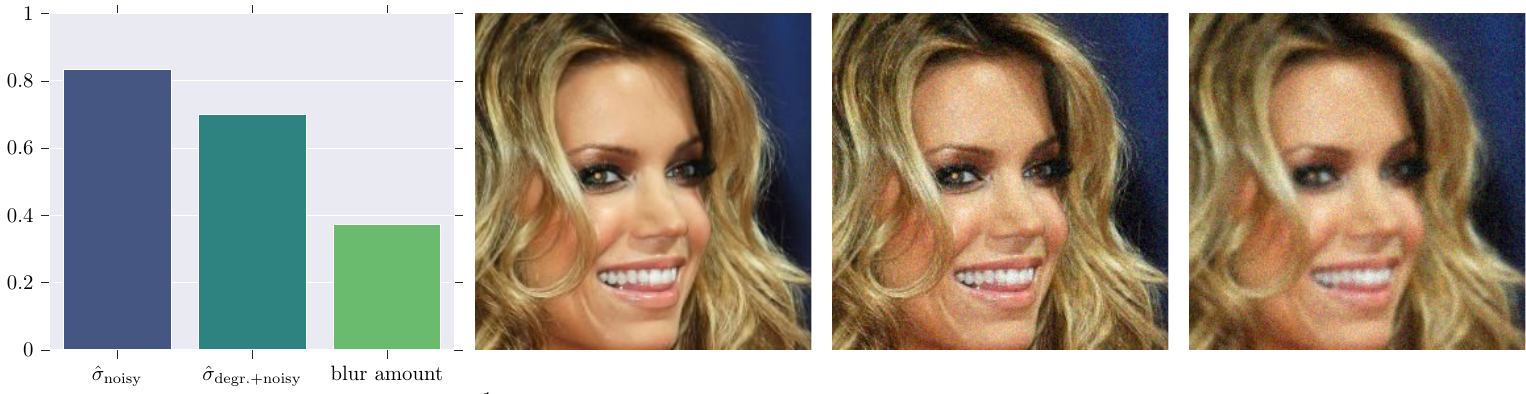}
	\end{subfigure}
	\caption{Contributors to severity. Degraded images with approx. the same $\hat{\sigma}$ may have different factors contributing to the predicted severity. The main contributor to $\hat{\sigma}$ in the top image is the image degradation (blur), whereas the bottom image is inherently more difficult to reconstruct.}\label{fig:contributors}
\end{figure}

\textbf{Identifying contributors to severity --} Finally, we analyze the contribution of different factors to the predicted degradation severities (Fig. \ref{fig:contributors}). We apply severity encoding to both the clean image with noise (no blur) and the noisy and degraded image, resulting in predicted severities $\hat{\sigma}_{noisy}$ and $\hat{\sigma}_{degr.+noisy}$. We quantify the difficulty of samples relative to each other via percentiles of the above two quantities, where we use $\hat{\sigma}_{noisy}$ as a proxy for the difficulty originating from the image structure. The reason for not using the clean image without additive noise is due to the fact that the severity encoder is trained on varying degradation severities, but with a fixed additive noise. Therefore, the model is not intended to accept noiseless measurements as inputs. We observe that for a fixed $\hat{\sigma}_{degr.+noisy}$, the composition of degradation severity may greatly vary. The two presented images have been assigned approximately the same  $\hat{\sigma}_{degr.+noisy}$, however the top image is inherently easy to encode (low $\hat{\sigma}_{noisy}$ percentile) compared to other images in the dataset, therefore the severity is mostly attributed to the image degradation. On the other hand, the bottom image with the same $\hat{\sigma}_{degr.+noisy}$ is less corrupted by blur, but with high $\hat{\sigma}_{noisy}$ indicating a difficult image. This example further corroborates the interaction between ground truth signal structure and the applied corruption in determining the difficulty of a reconstruction task. For a detailed discussion on the limitations of severity encoding, we refer the reader to Appendix \ref{apx:robustness}.

\begin{table*}
	\Huge
	\centering
	\renewcommand{\arraystretch}{1.3} 
	\resizebox{16.5cm}{!}{
		\begin{tabular}{ lccccccccccccccccccccccc }
			\toprule
			\textbf{FFHQ}&\multicolumn{4}{c}{\textbf{Gaussian Deblurring (Varying)}} & &\multicolumn{5}{c}{\textbf{Gaussian Deblurring (Fixed)}} & &\multicolumn{5}{c}{\textbf{Nonlinear Deblurring}}& &\multicolumn{5}{c}{\textbf{Random Inpainting}} \\
			\cmidrule{2-6}\cmidrule{8-12}\cmidrule{14-18}\cmidrule{20-24}
			\textbf{Method} &PSNR$(\uparrow)$&SSIM$(\uparrow)$&LPIPS$(\downarrow)$&FID$(\downarrow)$&$\overline{\text{NFE}}$&&PSNR$(\uparrow)$&SSIM$(\uparrow)$&LPIPS$(\downarrow)$&FID$(\downarrow)$&$\overline{\text{NFE}}$& &PSNR$(\uparrow)$&SSIM$(\uparrow)$&LPIPS$(\downarrow)$&FID$(\downarrow)$&$\overline{\text{NFE}}$& &PSNR$(\uparrow)$&SSIM$(\uparrow)$&LPIPS$(\downarrow)$&FID$(\downarrow)$&$\overline{\text{NFE}}$\\
			\midrule
			Latent-DPS & 23.69 & 0.6418 & 0.3579 & 87.26 & 1000 && 22.88 & 0.6136 & 0.3690 & 89.38 & 1000 && 22.07 & 0.5974 & 0.3814 & 90.89 & 1000 && 23.96 & 0.6566 & 0.3666 & 93.65 & 1000 \\
			\adaptive{Latent-DPS} & \underline{29.17} & \underline{0.8182} & \underline{0.2240} & \underline{55.57} & 100.3 && \underline{27.44} & \underline{0.7691} & \underline{0.2823} & \underline{80.44} & 127.7 && \underline{27.17} & \underline{0.7659} & \underline{0.2695} & \underline{69.78} & 136.1 && \underline{29.21} & \underline{0.8414} & \textbf{\underline{0.1945}} & \textbf{\underline{53.95}} & 104.7 \\
			\hline
			PSLD \citep{rout2023solving} & 25.06 &0.6769&0.3194&79.79&1000&&23.72  &0.6183&0.3324&88.45&1000&&- &-&-&-&- &&24.94&0.6617&0.3672&85.64&1000  \\
			\adaptive{PSLD} & \underline{29.26} & \underline{0.8205} & \textbf{\underline{0.2203}} &\textbf{ \underline{53.27}} & 100.3 && \underline{27.44} & \underline{0.7657} & \textbf{\underline{0.2797}} & \textbf{\underline{65.35}} & 127.7 && - & - & - & - & - && \underline{27.06} & \underline{0.8018} & \underline{0.2185} & \underline{55.12} & 104.7\\
			\hline
			GML-DPS \citep{rout2023solving}& 24.98 & 0.6884 & 0.3471 & 100.27 & 1000 &&  24.01&0.6574&0.3621&102.80&1000&& 23.00 & 0.6426 & 0.3812 & 108.79 & 1000 && 25.20 & 0.7044 & 0.3527 & 103.3 & 1000 \\
			\adaptive{GML-DPS} & \underline{29.21} & \underline{0.8276} & \underline{0.2274} & \underline{69.16} & 100.3 && \underline{27.47} & \underline{0.7699} & \underline{0.2816} & \underline{69.81} & 127.7 && \underline{27.11} & \underline{0.7640} & \underline{0.2756} & \underline{81.93} & 136.1 && \underline{28.95} & \underline{0.8437} & \underline{0.1957} & \underline{59.39} & 104.7 \\
			\hline
			ReSample \citep{song2023solving} & 28.77 & 0.8219 & 0.2587 & 81.96 & 500 && 27.62 & 0.7789 & 0.3148 & 102.47 & 500 && 26.61 & 0.7318 & 0.2838 & 68.57 & 500 && 27.51 & 0.7892 & 0.2460 & 63.39 & 500 \\
			\adaptive{ReSample} & \underline{29.07} & \underline{0.8330} & \underline{0.2383} & \underline{74.76} & 49.9 && \underline{27.77} & \underline{0.7845} & \underline{0.3092} & \underline{100.84} & 63.6 && \underline{26.88} & \underline{0.7660} &\textbf{\underline{0.2667}} & \textbf{\underline{64.57}} & 67.8 && \underline{28.13} & \underline{0.8260} & \underline{0.2045} & \underline{56.67} & 52.1 \\
			\hline
			\hline
			AE  &29.46&0.8358&0.2671&89.29&-&&27.69&0.7820&0.3396&110.56&-& &27.17&{0.7786}&0.3364 &111.24 &-& &{29.23}&0.8432&0.2515&85.87&-\\
			SwinIR \citep{liang_swinir_2021} &\textbf{30.69}&\textbf{0.8583}&0.2409&87.61&-&&\textbf{28.41}&\textbf{0.8021}&0.3091&108.49&-& &\textbf{27.60}&\textbf{0.7928}&0.3093&99.56&-& &\textbf{30.08}&\textbf{0.8654}&0.2223&78.32&- \\
			DPS \citep{chung2022diffusion}&	28.34&	0.7791&	0.2465&	81.70& 1000&&25.49&0.6829&0.3035&97.89& 1000&&	22.77&	0.6191&	0.3601&	109.58&1000& &28.30&0.8049&0.2451&82.78& 1000 \\
			\bottomrule
		\end{tabular}
	}
	\centering
	\renewcommand{\arraystretch}{1.3} 
	\resizebox{16.5cm}{!}{
		\begin{tabular}{ lccccccccccccccccccccccc }
			\addlinespace
			\addlinespace
			\toprule
			\textbf{LSUN Bedrooms}&\multicolumn{4}{c}{\textbf{Gaussian Deblurring (Varying)}} & &\multicolumn{5}{c}{\textbf{Gaussian Deblurring (Fixed)}} & &\multicolumn{5}{c}{\textbf{Nonlinear Deblurring}}& &\multicolumn{5}{c}{\textbf{Random Inpainting}} \\
			\cmidrule{2-6}\cmidrule{8-12}\cmidrule{14-18}\cmidrule{20-24}
			\textbf{Method} &PSNR$(\uparrow)$&SSIM$(\uparrow)$&LPIPS$(\downarrow)$&FID$(\downarrow)$&$\overline{\text{NFE}}$&&PSNR$(\uparrow)$&SSIM$(\uparrow)$&LPIPS$(\downarrow)$&FID$(\downarrow)$&$\overline{\text{NFE}}$& &PSNR$(\uparrow)$&SSIM$(\uparrow)$&LPIPS$(\downarrow)$&FID$(\downarrow)$&$\overline{\text{NFE}}$& &PSNR$(\uparrow)$&SSIM$(\uparrow)$&LPIPS$(\downarrow)$&FID$(\downarrow)$&$\overline{\text{NFE}}$\\
			\midrule
			Latent-DPS& 22.42 & 0.5984 & 0.4227 & 58.72 & 1000 && 21.53 & 0.5602 & 0.4391 & 58.34 & 1000 && 19.93 & 0.5430 & 0.4724 & 80.54 & 1000 &&22.29& 0.6056& 0.4355& 61.23& 1000 \\
			\adaptive{Latent-DPS} & \underline{28.05}& \underline{0.8099}& \underline{0.2185}& \underline{38.19}& 101.6&& \underline{25.92} & \underline{0.7410} & \underline{0.2892} & \underline{54.29} & 125.3 && \textbf{\underline{26.60}}& \underline{0.7709}& \underline{0.2672}& \underline{50.99}& 121.2&&\underline{27.73}& \underline{0.8472}& \textbf{\underline{0.1876}}& \underline{32.62}& 99.0 \\
			\hline
			PSLD \citep{rout2023solving} & 23.92 &0.6584&0.3427&52.40&1000&& 22.71 & 0.5942 & 0.3941 & 62.05 & 1000 && -  &-&-&-&-&&23.95 &0.6578&0.3808&55.25&1000  \\
			\adaptive{PSLD} & \underline{28.08} & \underline{0.8133} & \textbf{\underline{0.2146}} & \textbf{\underline{38.12}} & 101.6 && \underline{25.88} & \underline{0.7389} & \textbf{\underline{0.2872}} & \textbf{\underline{52.47}} & 125.3 && - & - & - & - & - && \underline{27.62} & \underline{0.8462} & \underline{0.1879} & \textbf{\underline{32.44}} & 99.0 \\
			\hline
			GML-DPS \citep{rout2023solving}& 23.55& 0.6508& 0.4253& 84.81& 1000& & 22.65 & 0.6173 & 0.4494 & 91.17 & 1000 && 21.75& 0.6038& 0.4591& 96.15& 1000&& 23.30& 0.6522& 0.4389& 88.49& 1000\\
			\adaptive{GML-DPS} & \underline{28.05}& \underline{0.8096}& \underline{0.2200}& \underline{38.16}& 101.6& & \textbf{\underline{25.94}} & \underline{0.7412} & \underline{0.2902} & \underline{54.27} & 125.3 && \underline{26.57}& \underline{0.7698}& \underline{0.2690}& \underline{50.97}& 121.2&& \textbf{\underline{27.82}}& \textbf{\underline{0.8498}}& \underline{0.1880}& \underline{34.25}& 99.0\\
			\hline
			ReSample \citep{song2023solving} & 28.19& 0.8193& 0.2499& 52.83& 500 & & \underline{25.84} & 0.7394 & 0.3401 & 83.30 & 500 && 26.20& 0.7658& 0.2709& 51.37& 500& & 27.17& 0.8121& 0.2235& 38.77& 500 \\
			\adaptive{ReSample} & \textbf{\underline{28.27}}& \textbf{\underline{0.8278}}& \underline{0.2284}& \underline{46.42}& 50.5& & 25.65 & \textbf{\underline{0.7426}} & \underline{0.3103} & \underline{65.59} & 62.4&& \underline{26.55}& \textbf{\underline{0.7837}}& \textbf{\underline{0.2586}}& \textbf{\underline{49.62}}& 60.3&& \underline{27.45}& \underline{0.8334}& \underline{0.1985}& \underline{32.71}& 49.2\\
			\bottomrule
		\end{tabular}
	}
	\caption{\label{tab:table_results} Experimental results on FFHQ (top) and LSUN Bedrooms (bottom). We \underline{underline} the better result between the baseline solver and its adapted variant, and use \textbf{bold} to highlight the best overall technique for the specific metric. Note that PSLD is only defined for linear inverse problems. Adaptive techniques consistently outperform their corresponding baseline across all metrics and achieve $8\times - 10\times$ speedup.}
\end{table*}

\subsection{Sample-adaptive reconstruction}
\begin{figure*}[ht]
	\centering
	\begin{subfigure}{.5\textwidth}
		\centering
		\includegraphics[width=0.99\linewidth]{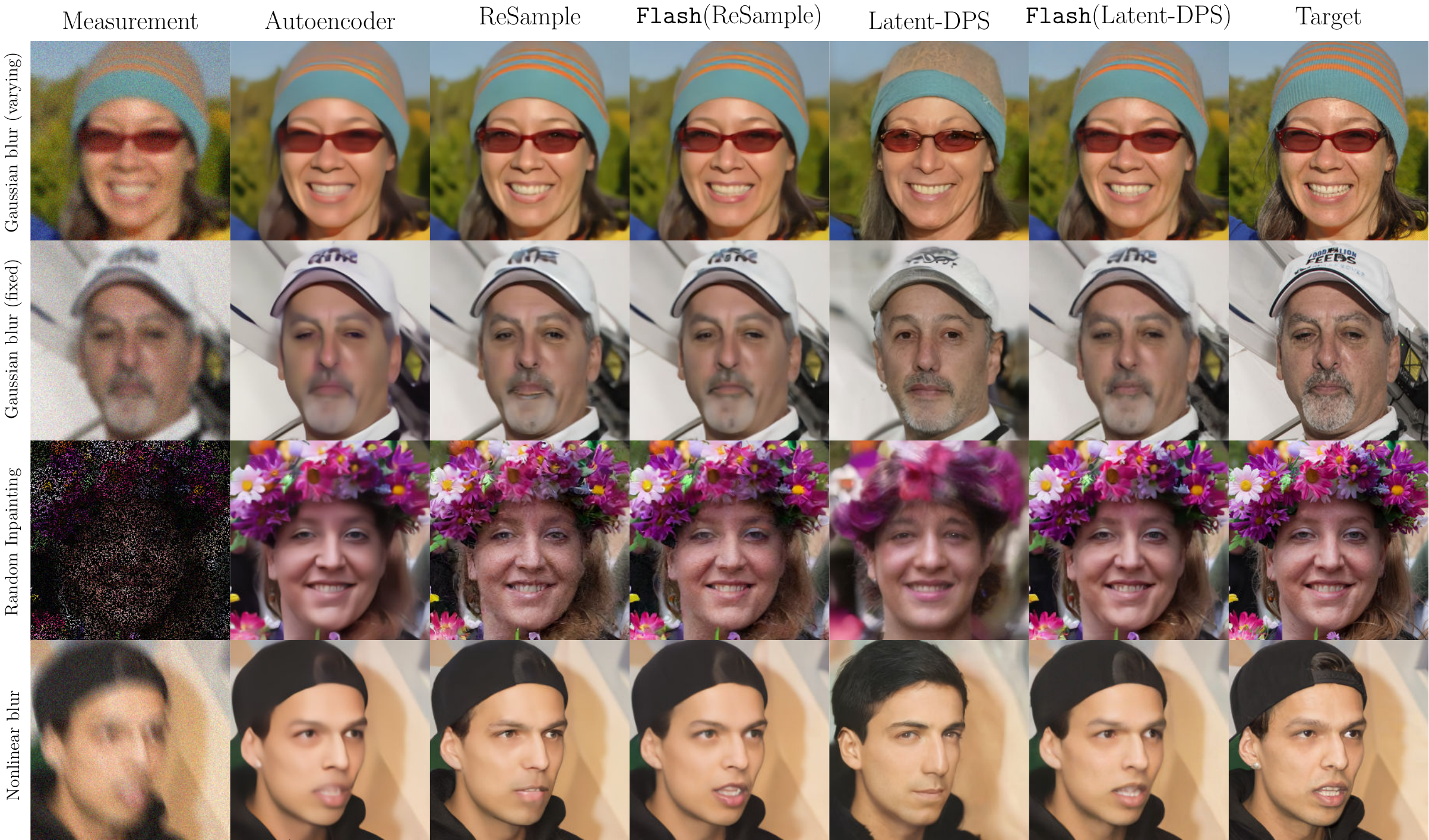}
	\end{subfigure}%
	\begin{subfigure}{.5\textwidth}
		\centering
		\includegraphics[width=0.97\linewidth]{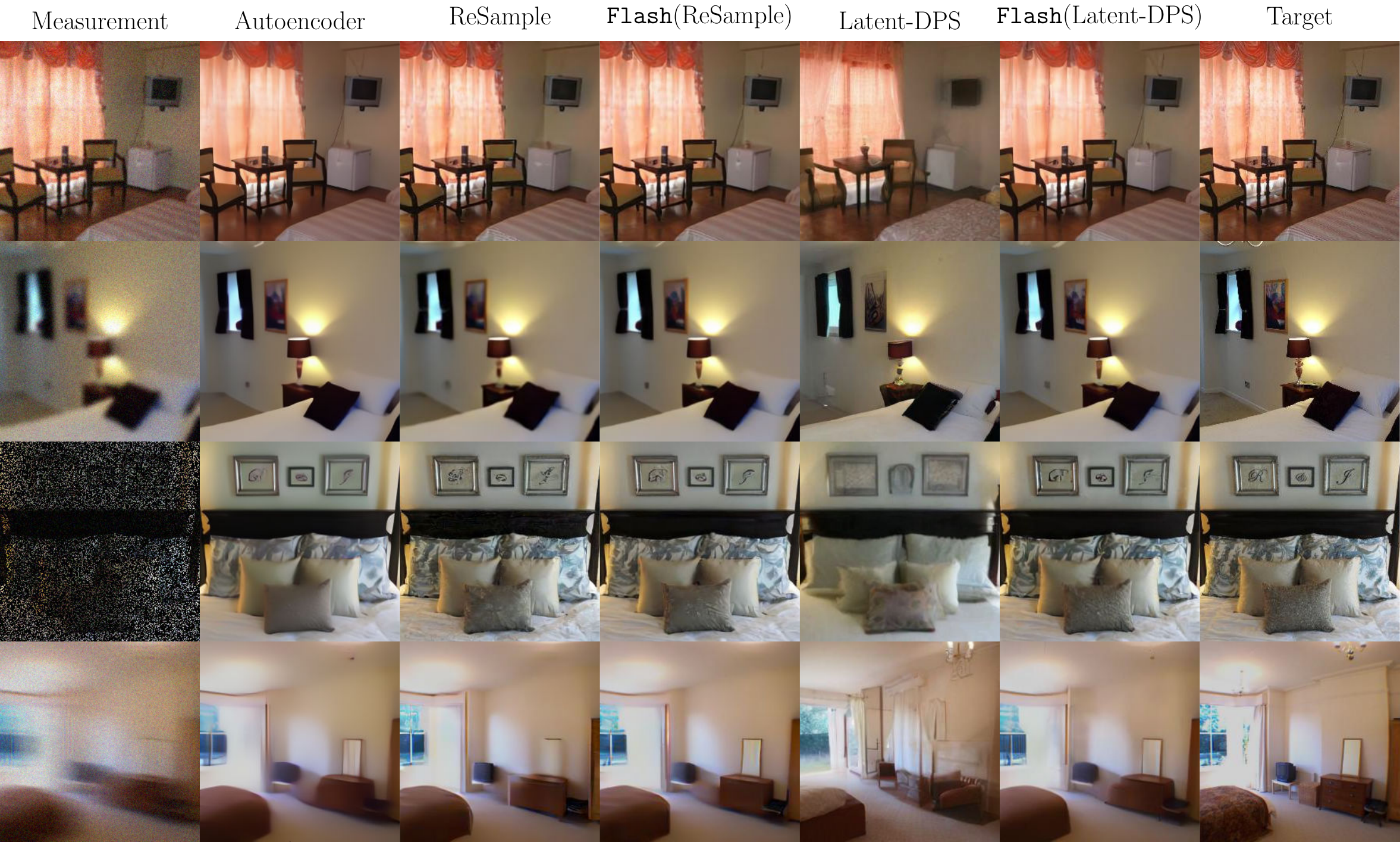}
	\end{subfigure}
	\caption{Visual comparison of reconstructions on the FFHQ (left) and LSUN Bedrooms (right) datasets.}
	\label{fig:visual_compare}
\end{figure*}

\textbf{Comparison of latent diffusion solvers--} Table \ref{tab:table_results} summarizes our experimental results comparing various latent diffusion solvers with and without our \methodname{} wrapper. Reconstructed samples are depicted in Figure \ref{fig:visual_compare}.  First, we observe that adaptive solvers consistently outperform their baselines across all degradations, datasets and image quality metrics. The performance improvement is especially significant in experiments with highly varying degradation severity, such as varying Gaussian blur, nonlinear blur with randomly sampled kernel and random inpainting, further highlighting the importance of sample-adaptivity. Interestingly, we observe that even though ReSample achieves the best performance among baseline solvers by a large margin, the performance gap diminishes between adaptive solvers with \adaptive{Latent-DPS} often outperforming more advanced techniques. We hypothesize that the differences between baseline solvers stem from differences in the diffusion dynamics in the early and middle stages (\textit{chaotic} and \textit{semantic} stages in \citet{song2023solving}) of sampling, a regime typically bypassed by \methodname{}. In terms of sampling speedup, we obtain $8\times - 10\times$ acceleration in average number of reverse diffusion steps across the test dataset compared to the non-adaptive baseline, without any drop in reconstruction quality.  Furthermore, we observe that simply leveraging DDIM acceleration in the baseline solvers, given the same compute budget in inference time, cannot compete with the reconstruction quality obtained from \methodname{} (details in Appendix \ref{apx:ddim}).

\begin{figure}[ht]
	\centering
	\begin{subfigure}{0.25\linewidth}
		\centering
		\includegraphics[width=0.99\textwidth]{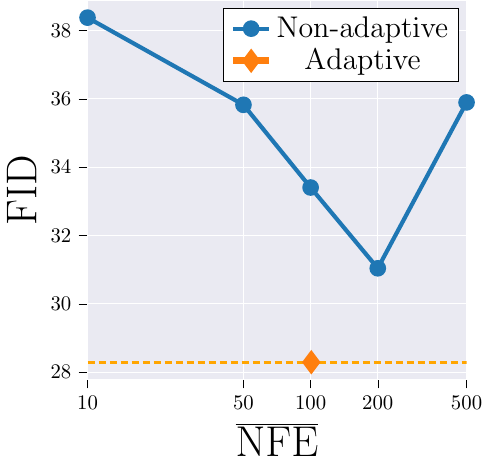}
	\end{subfigure}%
	\begin{subfigure}{0.25\linewidth}
		\centering
		\includegraphics[width=0.99\textwidth]{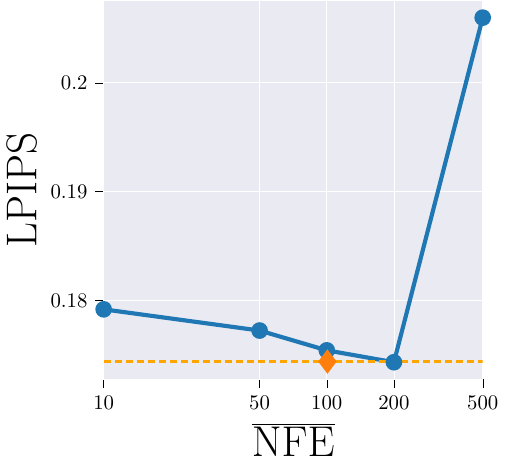}
	\end{subfigure}
	\begin{subfigure}{0.25\linewidth}
		\centering
		\includegraphics[width=0.99\textwidth]{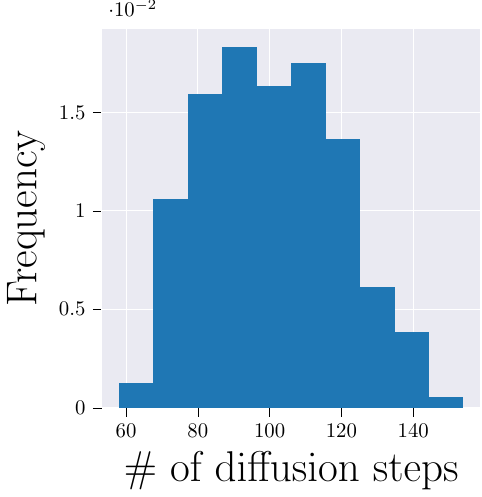}
	\end{subfigure}
	\caption[short]{Comparison of \adaptive{Latent-DPS} with a non-adaptive variant with fixed starting time (varying Gaussian blur, CelebA-HQ). \uline{Top}: Reconstruction quality as a function of the number of diffusion steps for the non-adaptive variant with various fixed starting times. The data point corresponding to our adaptive method depicts the average number of diffusion steps due to adaptive starting time. Our adaptive method achieves the best FID and near-optimal LPIPS compared to any choice of non-adaptive starting time. \uline{Bottom}: Histogram of predicted starting times for \adaptive{Latent-DPS}. \label{fig: nfe_efficiency}\vspace{-0.3cm}}
\end{figure}

\textbf{Comparison with other methods --}  In Table \ref{tab:table_results} (top), we observe that SwinIR, a state-of-the-art supervised restoration model achieves higher PSNR and SSIM than diffusion methods, however it falls short in terms of perceptual image quality. This phenomenon is due to the perception-distortion trade-off \citep{blau2018perception}: improving perceptual image quality is fundamentally at odds with distortion metrics. Furthermore, we highlight that our initial reconstructions obtained from the severity encoder are poorer quality than SwinIR reconstructions (compare AE vs. SwinIR), however diffusion samplers greatly improve over the perceptual quality of the initial reconstruction, eventually surpassing both SwinIR and all non-adaptive diffusion solver baselines.

\textbf{Importance of sample-adaptivity --} In order to highlight the importance of sample-adaptivity, we compare our adaptive technique to a simple non-adaptive modification. Here, instead of predicting the starting time via severity encoding on a sample-by-sample basis, we set the starting time for all samples to $N'$. For the sake of simplicity, we choose \adaptive{Latent-DPS} as our adaptive method, and vary the fixed start time between $10-500$ for the non-adaptive variant. We leverage SwinIR reconstructions as initialization for the non-adaptive variant. We perform experiments on CelebA-HQ under varying Gaussian blur. The results are depicted in Fig. \ref{fig: nfe_efficiency}. Our adaptive method achieves the best FID across any fixed number of steps by a large margin, despite the lower quality initialization from the severity encoder. Moreover, our technique with sample-adaptivity achieves near-optimal LPIPS with $2\times$ less average number of diffusion steps compared to the non-adaptive variant. We observe that the predicted start times from our technique are spread around the mean and not closely concentrated, further highlighting the adaptivity of \methodname{}. 

\textbf{Robustness and limitations--} The performance of \methodname{} relies on the accuracy of the severity encoder trained in a supervised fashion on the degraded image distribution. Therefore, it is crucial to understand the robustness of severity encoding to noise level and forward model shifts in test time. In terms of noise, we observe that severity encoding maintains its performance on noise levels lower than in the training setting, however it breaks down at noise levels well above the training regime. In terms of forward model mismatch, we find that as long as the train and test operators are similar (heavy Gaussian blur and nonlinear blur), severity encoding can maintain acceptable performance. However in case of significant mismatch, the accuracy of severity encoding breaks down. An in-depth characterization of severity encoder robustness is detailed in Appendix \ref{apx:robustness}.

\begin{figure}[ht]
	\centering
	\includegraphics[width=0.7\linewidth]{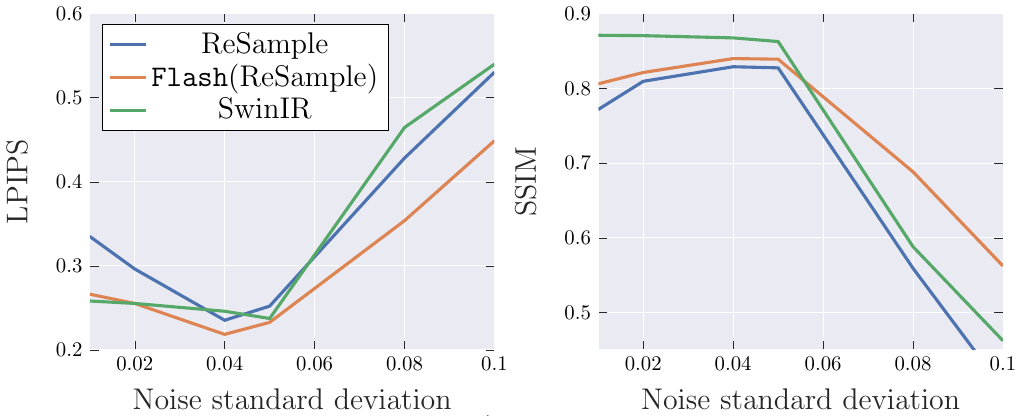}
	\caption{Robustness experiment: we perturb the forward model noise level and compare the performance of ReSample and its adaptive variant. We observe improved robustness after imbuing the solver with sample-adaptivity.\vspace{-0.3cm}}
	\label{fig:robustness}
\end{figure}

Moreover, we analyze the robustness of \methodname{} solvers compared to their non-adaptive counterparts with respect to a mismatch in forward model noise level. In particular, we base our comparison on ReSample, the best performing baseline solver in our experiments. Fig. \ref{fig:robustness} demonstrates that the performance of ReSample degrades more rapidly than its adaptive counterpart when the noise level is higher than expected, following a similar trend to SwinIR, a supervised non-diffusion reconstructor. In the low-noise regime, \adaptive{ReSample} shows improved robustness compared to the baseline solver. More experiments on forward model mismatch can be found in Appendix  \ref{apx:fwd_robustness}.
\section{Conclusions}
In this work, we make the key observation that the difficulty of solving an inverse problem may vary greatly on a sample-by-sample basis, depending on the ground truth signal structure, the applied corruption, the model, the training set and the complex interactions between these factors. Despite this natural variation in the difficulty of a reconstruction task,  most techniques apply a rigid model that expends the same compute irrespective of the amount of information destroyed in the noisy measurements, resulting in suboptimal performance and wasteful resource allocation. We propose \methodname{}, a sample-adaptive method that predicts the degradation severity of corrupted signals, and utilizes this estimate to automatically tune the compute allocated to reconstruct the sample. In particular, we use the prediction error of an encoder in latent space as a proxy for reconstruction difficulty, which we call severity encoding. Then, we leverage a latent diffusion process to reconstruct the sample, where the length of the sampling trajectory is directly scaled by the predicted severity. Our method can be coupled with any latent diffusion solver, greatly enhancing its performance and accelerating sampling by up to $10\times$. We demonstrate the gains in terms of reconstruction quality and speedup on a variety of degradations and datasets.
\section*{Impact Statement}
The presented paper leverages deep learning for reconstructing corrupted images. Deep learning-based reconstruction approaches are known to hallucinate realistic-looking features that may be inconsistent with the ground truth. Therefore, deploying our technique in safety-critical applications requires additional caution. Moreover, the technique leverages diffusion models capable of generating photorealistic synthetic images, which in turn can be abused by bad actors to spread misinformation.
\section*{Acknowledgments}
We thank the anonymous reviewers who provided valuable suggestions and constructive feedback on our work.  M. Soltanolkotabi is supported by the Packard Fellowship in Science and Engineering, a Sloan
Research Fellowship in Mathematics, an NSF-CAREER under award \#1846369, DARPA FastNICS program, and NSF-CIF awards \#1813877 and \#2008443. and NIH DP2LM014564-01. This research is also in part supported by AWS credits through an Amazon Faculty research award.
\bibliography{references}
\bibliographystyle{icml2024}

\newpage
\appendix
\onecolumn
\section{Experimental Details}\label{apx:experimental_details}
Here we provide additional details on the experimental setup and hyperparameters.

\textbf{Datasets --} For experiments on CelebA-HQ and FFHQ datasets, we use the train and validation splits provided in the GitHub repo of "Taming Transformers"\footnote{\url{https://github.com/CompVis/taming-transformers/tree/master/data}}. For LSUN Bedrooms experiments, we use the custom split provided in GitHub repo of "Latent Diffusion"\footnote{\label{foot:ldm} \url{https://github.com/CompVis/latent-diffusion}}. For the test split, we subsample $200$ ids from the corresponding validation ids files. Specific ids we used will be available when the codebase is released. We provide the specifics for each comparison method next.

\textbf{Model architecture --} In all experiments, we use LDM models pre-trained on the CelebA-HQ and LSUN Bedrooms datasets out of the box. We fine-tune the severity encoder from the LDM's pre-trained encoder. We obtain the degradation severity estimate $\hat{\sigma} \in \mathbb{R}^+$ from the latent reconstruction $\hat{\vct{z}} \in \mathbb{R}^d$ as 
\vspace{-0.2cm}
\begin{equation*}
	\hat{\sigma} = \frac{1}{d}\sum_{i=1}^d [Conv(\hat{\vct{z}})^2]_i,
\end{equation*}
where $Conv(\cdot)$ is a learned $1\times1$ convolution with $d$ input and $d$ output channels.

\textbf{Training setup -- } We train severity encoders using Adam optimizer with batch size $28$ and learning rate $0.0001$ for about $200k$ steps until the loss on the validation set converges. We use Quadro RTX 5000 and Titan GPUs.

\textbf{Hyperparameters --} We scale the reconstruction loss terms with their corresponding dimension ($d$ for $L_{lat. rec.}$ and $n$ for $L_{im. rec.}$), which we find to be sufficient without tuning for $\lambda_{im. rec.}$. We tune $\lambda_\sigma$ via grid search on $[0.1, 1, 10]$ on the varying Gaussian blur task and set to $10$ for all experiments. 


For \texttt{Flash} experiments, we tune the noise correction parameter $c$ on the validation subset by grid search over $[0.8, 1.0, 1.2]$ after tuning the remaining baseline method hyperparameters.

Next, we provide details about baseline solvers, comparison methods and hyperparameter tuning.

\textbf{DPS}  \citep{chung2022diffusion}: DPS approximates the gradient of the likelihood as 
$$\nabla_{\vct{x}_i}  \log p(\vct{y} | \vct{x}_i) \approx \nabla_{\vct{x}_i}  \log p(\vct{y} | \hat{\vct{x}}_0(\vct{x}_i)),$$
where $\hat{\vct{x}}_0(\vct{x}_i)$ is the posterior mean estimate obtained from the score model, resulting in the update rule 
\begin{equation}
	\vct{x}_{i-1} = \vct{x'}_{i-1} - \eta_t \nabla_{\vct{x}_i} \left\| \fwd{\hat{\vct{x}}_0(\vct{x}_i} - \vct{y} \right\|^2,
\end{equation}
where $ \vct{x'}_{i-1}$ is the result of the unconditional diffusion step and  
\begin{equation}\label{apx:eta}
	\eta_i^{DPS} = \frac{\eta^{DPS}}{ \left\| \fwd{\hat{\vct{x}}_0(\vct{x}_i)} - \vct{y} \right\|},
\end{equation}
with $\eta^{DPS} > 0$ tuning parameter. 

We use the code implementation in the official GitHub repository\footnote{\url{https://github.com/DPS2022/diffusion-posterior-sampling}}. For all other solvers, we base our implementation on the LDM repository$^{\ref{foot:ldm}}$. We tune $\eta^{DPS} > 0$ by performing a grid search over $[0.1, 0.5, 1.0 ,2.0, 3.0, 5.0, 10.0]$ for all experiments. We provide the optimal hyperparameters in Table \ref{tab:hyperparams}.

\textbf{Latent-DPS}: This solver is the direct application of DPS in the latent space of an autoencoder that uses the approximation
\begin{equation}\label{eq:ldps_approx}
\nabla_{\vct{z}_i}  \log p(\vct{y} | \vct{z}_i) \approx \nabla_{\vct{z}_i}  \log p(\vct{y} | \mathcal{D}_0(\hat{\vct{z}}_0(\vct{z}_i))),
\end{equation}
leading to the updates 
\begin{equation}
	\vct{z}_{i-1} = \vct{z'}_{i-1} - \eta_i \nabla_{\vct{z}_i} \left\| \fwd{\dec{\hat{\vct{z}}_0(\vct{z}_i}} - \vct{y} \right\|^2.
\end{equation}
Here $ \vct{z'}_{i-1}$ is the result of the unconditional latent diffusion update and  
\begin{equation}\label{apx:eta_ldps}
	\eta_i^{LDPS} = \frac{\eta^{LDPS}}{ \left\| \fwd{\dec{\hat{\vct{z}}_0(\vct{z}_t)}} - \vct{y} \right\|}
\end{equation}
with $\eta^{LDPS} > 0$ tuning parameter. We tune $\eta^{LDPS} > 0$ by performing a grid search over $[0.1, 0.2, 0.3, 0.4, 0.5, 1.0]$ for all baseline FFHQ experiments, and over $[0.1, 0.2, 0.4, 1.0]$ for baseline LSUN Bedrooms experiments. For \adaptive{Latent-DPS} we observe larger $\eta^{LDPS}$ values work better. Hence, we search over $[1.0,2.0,3.0,4.0,7.0,10.0]$ for FFHQ experiments and $[1.0,2.0,4.0,10.0]$ for LSUN Bedrooms experiments. We provide the optimal hyperparameters in Table \ref{tab:hyperparams}.

\textbf{GML-DPS} \citep{rout2023solving}: Latent-DPS achieves poor performance out-of-the-box due to the inaccuracy of the approximation in \ref{eq:ldps_approx} in early stages of diffusion and due to additional complications introduced by the decoder \cite{rout2023solving, song2023solving, chung2023prompt}. GML-DPS is an extension of Latent-DPS that guides the diffusion towards fixed points of the encoder-decoder composition via the "goodness modified" approximation
\begin{equation}\label{eq:gmldps_approx}
	\nabla_{\vct{z}_i}  \log p(\vct{y} | \vct{z}_i) \approx \nabla_{\vct{z}_i}  \log p(\vct{y} | \mathcal{D}_0(\hat{\vct{z}}_0(\vct{z}_i))) + \lambda \nabla_{\vct{z}_i} \| \hat{\vct{z}}_0(\vct{z}_i) -  \enc{\mathcal{D}_0(\hat{\vct{z}}_0(\vct{z}_i))}\|^2,
\end{equation}
with some $\lambda >0$ weight. We normalize the additional guidance term similarly to DPS resulting in the update 
\begin{equation}
	\vct{z}_{i-1} = \vct{z'}_{i-1} - \eta_i^{LDPS} \nabla_{\vct{z}_i} \left\| \fwd{\dec{\hat{\vct{z}}_0(\vct{z}_i)}} - \vct{y} \right\|^2 - \gamma_i^{GM} \nabla_{\vct{z}_i} \| \hat{\vct{z}}_0(\vct{z}_i) -  \enc{\mathcal{D}_0(\hat{\vct{z}}_0(\vct{z}_i))}\|^2,
\end{equation}
where 
\begin{equation}\label{apx:gamma_gm}
	\gamma_i^{GM} = \frac{\gamma^{GM}}{ \left\| \hat{\vct{z}}_0(\vct{z}_i) -  \enc{\mathcal{D}_0(\hat{\vct{z}}_0(\vct{z}_i))}\right\|},
\end{equation}
and $\gamma^{GM}$ is a hyperparameter to be tuned. We tune $\eta^{LDPS} > 0$ and $\gamma^{GM} > 0$ by performing a greedy 2D grid search (first tune $\eta^{LDPS}$ with $\gamma^{GM}=0.1$ and then tune $\gamma^{GM}$ with the best $\eta^{LDPS}$ value) over $[0.1, 0.2, 0.3, 0.4, 0.5, 1.0] \times [0.05, 0.1, 0.2, 0.5]$ for all baseline FFHQ experiments, and over $[0.1, 0.2, 0.4, 1.0] \times [0.05, 0.1, 0.2, 0.5]$ for baseline LSUN Bedrooms experiments. For \adaptive{GML-DPS}, we search over $[1.0,2.0,3.0,5.0,10.0] \times [0.05,0.1,0.2,0.5]$ for FFHQ experiments and $[1.0,2.0,4.0,10.0] \times [0.05,0.1,0.2,0.5]$ for LSUN Bedrooms experiments. We provide the optimal hyperparameters in Table \ref{tab:hyperparams}.

\textbf{PSLD} \citep{rout2023solving}: PSLD improves upon GML-DPS by adding a "gluing" term in order to avoid inconsistencies at mask boundaries. The approximation takes the form
\begin{equation}\label{eq:psld_approx}
	\nabla_{\vct{z}_i}  \log p(\vct{y} | \vct{z}_i) \approx \nabla_{\vct{z}_i}  \log p(\vct{y} | \mathcal{D}_0(\hat{\vct{z}}_0(\vct{z}_i))) + \lambda \nabla_{\vct{z}_i} \| \hat{\vct{z}}_0(\vct{z}_i) -  \enc{ \mathcal{A}\vct{x}_0 + (\mathbf{I} - \mathcal{A} \mathcal{A}^T)   \dec{\hat{\vct{z}}_0(\vct{z}_i)}}\|^2,
\end{equation}
leading to updates in the form
\begin{equation}
	\vct{z}_{i-1} = \vct{z'}_{i-1} - \eta_i^{LDPS} \nabla_{\vct{z}_i} \left\| \fwd{\dec{\hat{\vct{z}}_0(\vct{z}_i)}} - \vct{y} \right\|^2 - \gamma_i^{PSLD} \nabla_{\vct{z}_i} \| \hat{\vct{z}}_0(\vct{z}_i) -  \enc{  \mathcal{A}\vct{x}_0+ (\mathbf{I} - \mathcal{A} \mathcal{A}^T)   \dec{\hat{\vct{z}}_0(\vct{z}_i)}}\|^2,
\end{equation}
where $\vct{x}^*$ is the ground truth clean signal and
\begin{equation}\label{apx:gamma_psld}
	\gamma_i^{PSLD} = \frac{\gamma^{PSLD}}{ \left\| \hat{\vct{z}}_0(\vct{z}_i) - \enc{ \mathcal{A}\vct{x}_0 + (\mathbf{I} - \mathcal{A} \mathcal{A}^T)   \dec{\hat{\vct{z}}_0(\vct{z}_i)}}\right\|},
\end{equation}
and $\gamma^{PSLD}$ is a hyperparameter. Note that PSLD requires access to noiseless measurements $\mathcal{A}\vct{x}_0$ and to the transposed of the linear operator $\mathcal{A}$. In our experiments that involve measurement noise, we replace $\mathcal{A}\vct{x}_0$  with $\vct{y}$ (noisy measurement). For Gaussian blur, we implement the transposed operator $\mathcal{A}^T$ using \texttt{ConvTranspose2d} in \texttt{torch}. We omit experiments involving nonlinear blur with PSLD as the transposed operator is not known. 

We tune $\eta^{LDPS} > 0$ and $\gamma^{PSLD} > 0$ by performing a greedy 2D grid search over $[0.1, 0.2, 0.3, 0.4, 0.5, 1.0] \times [0.05, 0.1, 0.2, 0.5]$ for all baseline FFHQ experiments, and over $[0.1, 0.2, 0.4, 1.0] \times [0.05, 0.1, 0.2, 0.5]$ for baseline LSUN Bedrooms experiments. For \adaptive{PSLD}, we search over $[1.0,2.0,4.0,10.0] \times [0.05,0.1,0.2,0.5]$ for all experiments. We provide the optimal hyperparameters in Table \ref{tab:hyperparams}.

\textbf{ReSample} \citep{song2023solving}: Instead of DPS-like approximations as in the previous solvers, ReSample applies data consistency directly on the posterior mean estimates by solving the latent-domain optimization problem
 \begin{equation}\label{apx:eq_hc}
\hat{\vct{z}}_0(\vct{y}) = \text{arg}\min_{\vct{z}} \| \vct{y} - \fwd{\mathcal{D}_0(\vct{z})}\|_2^2 + \lambda \|\vct{z} -\hat{\vct{z}}_0(\vct{z}_i) \|_2^2,
 \end{equation} 
 or its pixel-domain variant  after passing $\hat{\vct{z}}_0(\vct{z}_i)$ through the decoder. To solve the above problem, numerical methods are used where optimization is terminated when the stopping condition $ \frac{1}{m}\|\vct{y} - \fwd{\mathcal{D}_0(\vct{z})}\|_2^2 < \tau$ is reached. $\tau$ is a hyperparameter, which we set to $\sigma_y^2$ as reducing the error below the noise level amounts to overfitting to the noisy measurement. Furthermore, we set $\lambda=1$ for all experiments following \cite{song2023solving}, as we observe that the method is not sensitive to the choice of this parameter as long as the problem is solved to the desired accuracy. We use Adam optimizer with step size $0.1$ and initial guess $\hat{\vct{z}}_0(\vct{z}_i)$ to minimize the objective in \eqref{apx:eq_hc}. As solving \eqref{apx:eq_hc} is computationally heavy, it is only performed every $K$ steps, where we set $K=10$ for all experiments following the setting in the original paper.
 
 To map  $\hat{\vct{z}}_0(\vct{y})$ back to the noisy latent manifold, authors introduce stochastic resampling in the form
 \begin{equation}\label{apx:eq_stochastic_resamp}
 	\vct{z}_i = \frac{\omega_i^2 \sqrt{\bar{\alpha}_i} \hat{\vct{z}}_0(\vct{y}) + (1 - \bar{\alpha}_i)\vct{z}_i'}{\omega_i^2 + (1 - \bar{\alpha}_i)} + \sqrt{\frac{\omega_i^2 (1 - \bar{\alpha}_i)}{\omega^2_i + (1-\bar{\alpha}_i)}} \epsilon, ~~ \epsilon \sim \mathcal{N}(\mathbf{0}, \mathbf{I}),
 \end{equation}
 where $\omega_i$ controls the trade-off between prior consistency and data consistency and is a hyperparameter. Authors choose the adaptive schedule 
 \begin{equation}
 	\omega_i^2 = \omega \left(\frac{1 - \bar{\alpha}_{i-1}}{\bar{\alpha}_i}\right) \left(1 - \frac{\bar{\alpha}_i}{\bar{\alpha}_{i-1}}\right),
 	\end{equation}
 where only $\omega$ needs to be tuned. We do not use time-travel updates as in the original algorithm as it appears to have only minor contribution to performance (see Table 8. in \citet{song2023solving}). Furthermore, we use DDIM sampling with $N=500$ matching the setting in the original paper.
 
 We tune $\omega > 0$ by performing a grid search over $[0, 10, 20 ,40, 100, 200, 400, 1000, 2000, 4000]$ for all experiments. We provide the optimal hyperparameters in Table \ref{tab:hyperparams}.
 
\textbf{SwinIR: } For  all experiments, we train SwinIR using Adam optimizer with batch size $28$ for $100$ epochs. We use learning rate $0.0002$ for the first $90$ epochs and drop it by a factor of $10$ for the remaining $10$ epochs. We use Quadro RTX 5000 and Titan GPUs. We use $4$ RSTB blocks, each with $6$ STLs and embedding dimension $96$.

\textbf{Autoencoded (AE): } We use the latent at severity encoders output and decode it without reverse diffusion to get the reconstruction.

We provide all the optimal hyperparameters we use in our experiments in Table \ref{tab:hyperparams}.
\begin{table}[H]
	\huge
	\centering
	\renewcommand{\arraystretch}{1.5} 
	\resizebox{17.0cm}{!}{
	\begin{tabular}{|c|c|c|c|cc|cc|ccc|cc|ccc|c|cc|}
		\toprule
	&& DPS & Latent-DPS & \multicolumn{2}{|c|}{\adaptive{Latent-DPS}} & \multicolumn{2}{|c|}{GML-DPS} & \multicolumn{3}{|c|}{\adaptive{GML-DPS}}  & \multicolumn{2}{|c|}{PSLD}      & \multicolumn{3}{|c|}{\adaptive{PSLD}}  & ReSample & \multicolumn{2}{|c|}{\adaptive{ReSample}} \\
		\cline{3-19} 
			\textbf{Dataset}&\textbf{Operator}& $\eta^{DPS}$ & $\eta^{LDPS}$ & $\eta^{LDPS}$& $c$&$\eta^{LDPS}$&$\gamma^{GM}$&$\eta^{LDPS}$&$\gamma^{GM}$& $c$ & $\eta^{LDPS}$&$\gamma^{PSLD}$ & $\eta^{LDPS}$&$\gamma^{PSLD}$ & $c$ & $\omega$ & $\omega$ & $c$   \\
		 \hline
		\multirow{4}{*}{FFHQ}          & Gaussian blur (varying) & 5.0 & 0.1 & 1.0 &1.2 & 0.2 & 0.05 & 3.0&0.05&1.2 & 0.2&0.05 & 2.0&0.2&1.0 & 400  & 400&1.2  \\
		& Gaussian blur (fixed) & 3.0 & 0.1  & 3.0&1.2   & 0.2 &0.05 & 2.0&0.05&1.0  & 0.2&0.1  & 2.0&0.1&1.0    & 4000     & 2000&1.2      \\ 
		& Nonlinear blur                 & 0.5 & 0.1  & 1.0&1.0   & 0.2&0.1  & 3.0&0.05&1.2  & -          & -      & - & -    &-      & 400      & 200&1.0       \\ 
		& Random inpainting              & 3.0 & 0.3  & 7.0&1.2   & 0.5&0.1  & 10.0&0.05&1.2 & 0.5&0.05 & 10.0& 0.5& 1.0 & 200      & 200&1.2       \\ \hline
		\multirow{4}{*}{LSUN Bedrooms} & Gaussian blur (varying) & -   & 0.1 & 2.0&1.2 & 0.2&0.1  & 2.0&0.05&1.2 & 0.2&0.2  & 2.0&0.2&1.0 & 2000 & 1000&1.2 \\ 
		& Gaussian blur (fixed) & -   & 0.1  & 2.0&1.0   & 0.2&0.1  & 2.0&0.05&1.0  & 0.2&0.1  & 2.0&0.1&1.0    & 2000     & 200&1.2       \\ 
		& Nonlinear blur                 & -   & 0.1  & 2.0&1.0   & 0.2&0.1  & 2.0&0.05&1.0  & -          & -    & - & -     &-       & 1000     & 1000&1.0      \\ 
		& Random inpainting              & -   & 0.2  & 10.0&1.2  & 0.4&0.1  & 10.0&0.05&1.0 & 0.4&0.05 & 10.0&0.05&1.2  & 1000     & 400&1.2       \\ \hline
	\end{tabular}
} \caption{\label{tab:hyperparams} Optimal hyperparameters used in \methodname{} experiments.}
\end{table}

%

\section{Ablation of \methodname{} Components}
\methodname{} combines sample-by-sample adaptation of the sampling trajectory via severity encoding with latent-space diffusion. We ablate the effect of these two key components and demonstrate their relative contribution to the performance of \methodname{}. In particular, we choose \adaptive{Latent-DPS} as our adaptive technique and we ablate the effect of adaptation by fixing the starting time for each sample to the average predicted starting time of  \adaptive{Latent-DPS} on the corresponding task ($N=100$ for Gaussian deblurring, $N = 134$ for nonlinear deblurring). We further ablate the effect of latent domain reconstruction via repeating the above experiment with a pixel-space score model. We optimized all models for best performance in terms of LPIPS via grid search over $\eta=[0.5, 1.0, 1.5, 2.0, 3.0, 5.0, 20.0]$. Our findings are summarized in Table \ref{tab:component_ablation}. We observe, that adaptation via severity encoding provides a significant boost to the performance in terms of LPIPS. When comparing non-adaptive methods, we obtain similar results with and without latent space diffusion, however latent diffusion is preferable since (1) it has greatly reduced computational cost and (2) degradation severity is better captured in latent space.
 
\begin{table}[htp]
	\centering
	\resizebox{14cm}{!}{
		\begin{tabular}[H]{ lcccccccccc }
			\toprule
			& &\multicolumn{4}{c}{\textbf{Gaussian Deblurring}} & &\multicolumn{4}{c}{\textbf{Nonlinear Deblurring}} \\
			\cmidrule{3-6}\cmidrule{8-11}
			\textbf{Adaptive?}& \textbf{Latent?} &PSNR$(\uparrow)$&SSIM$(\uparrow)$&LPIPS$(\downarrow)$&FID$(\downarrow)$& &PSNR$(\uparrow)$&SSIM$(\uparrow)$&LPIPS$(\downarrow)$&FID$(\downarrow)$ \\
			\midrule
			\checkmark& \checkmark&	29.16&\underline{0.8191}&\textbf{0.2241}&	\textbf{29.46}& &	\underline{27.22 }&	\underline{0.7705}&	\textbf{0.2694}&		\textbf{36.92} \\
			\hline
			\xmark& \checkmark&	\textbf{29.55}&	\textbf{0.8377}&	\underline{0.2346}&	\underline{49.06}& &	\textbf{27.35}&	\textbf{0.7844}&	{0.2847}&\underline{54.25} \\
			\hline
			\xmark& \xmark&	\underline{29.23}&	{0.8058}&	{0.2377}&	{54.20}& &	{26.83}&	{0.7379}&	\underline{0.2843}&{57.82} \\
			\bottomrule 
		\end{tabular}
	}
	\caption{Ablation of different components of the \methodname{} framework on the FFHQ dataset. \label{tab:component_ablation}}
\end{table}

\section{Robustness Studies}\label{apx:robustness}
\subsection{Noise robustness}\label{apx:noise_robustness}
\textbf{Robustness of severity encoding --} As demonstrated throughout the paper, our severity encoder provides useful estimates of the degradation severity of samples in the scenario where the degradation and noise level matches that in the training setup. However, it is very important to understand the limitations of severity encoding in the presence of test-time shifts in the corruption process.

Our most crucial expectation towards the severity encoder is to predict a degradation severity $\hat{\sigma}(\vct{y})$ that is a non-decreasing function of the true underlying corruption level. More specifically, assume that the forward model $\mathcal{A}$ is parameterized by a severity level $s\in [0, 1]$, with $s=0$ being least severe, $s=1$ being most severe. For instance, the standard deviation of the Gaussian blur kernel may serve as $s$ in case of Gaussian blur (see a more rigorous treatment of degradation severity in \citet{fabian2023diracdiffusion}). Then for a given specific clean image $\vct{x}_0$ and its corrupted versions $\vct{y}_{s'} := \mathcal{A}(\vct{x}_0; s') + \vct{n}, ~~\vct{y}_{s''} : = \mathcal{A}(\vct{x}_0; s'') + \vct{n}$, we expect our severity encoder to satisfy
\begin{equation}\label{eq:ordering_req}
\hat{\sigma}(\vct{y_{s'} }) \leq \hat{\sigma}(\vct{y_{s''} }), ~~~, ~~\forall s', s'' \in [0, 1] , ~ s' < s'',
\end{equation}
which we call the \textit{monotonicity requirement}, i.e. the ordering of predicted severities need to match the ordering of the true underlying severity levels. 

\begin{figure}[htp]
	\centering
	\begin{subfigure}[b]{0.72\textwidth}
		\hspace{-0.3cm}
		\includegraphics[width=1\linewidth]{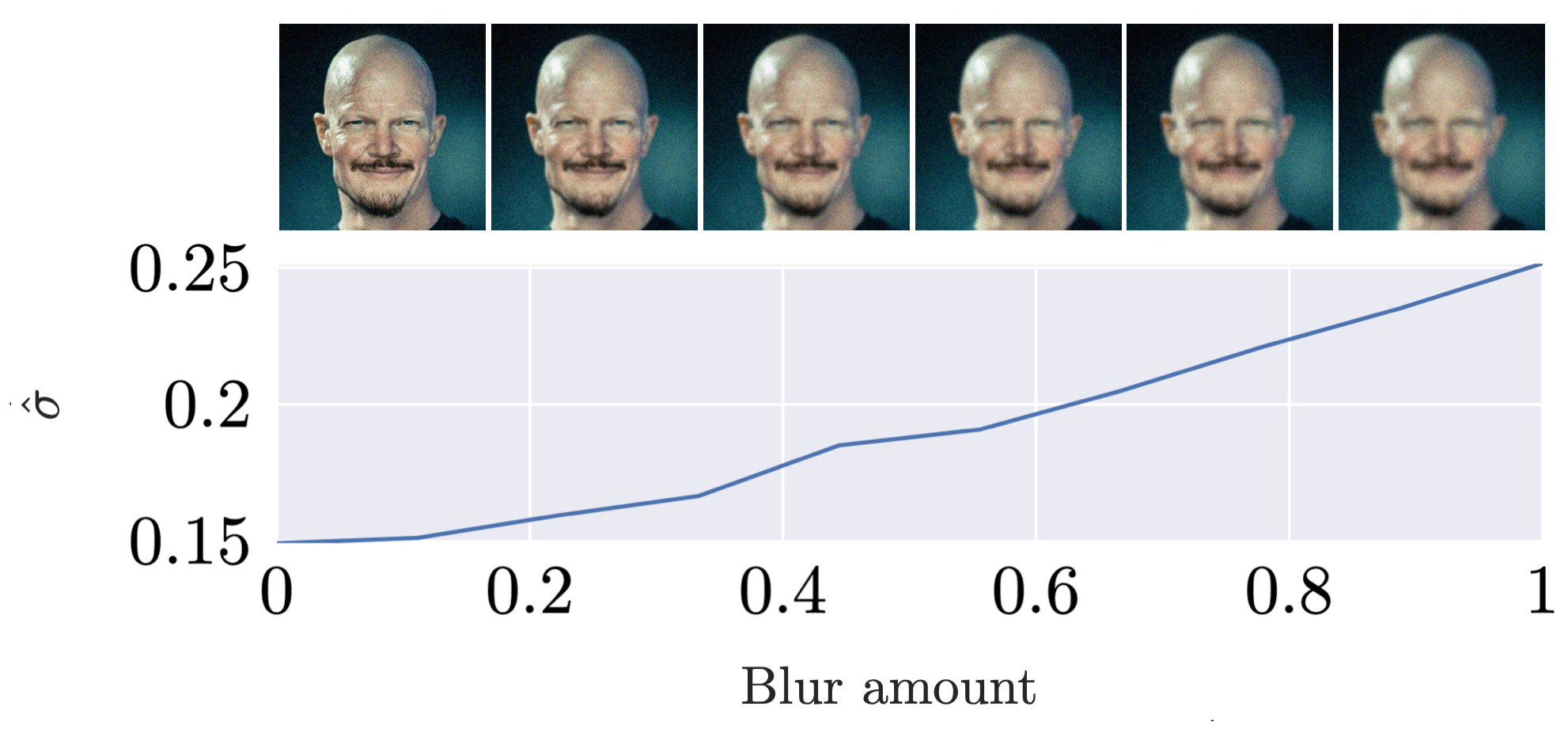}
		\caption{Measurement noise: $\sigma_{\vct{y}} = 0.05$ (no test-time perturbation).}
		\label{fig:ordering_correct} 
	\end{subfigure}
	\centering
	\begin{subfigure}[b]{0.7\textwidth}
		\includegraphics[width=1\linewidth]{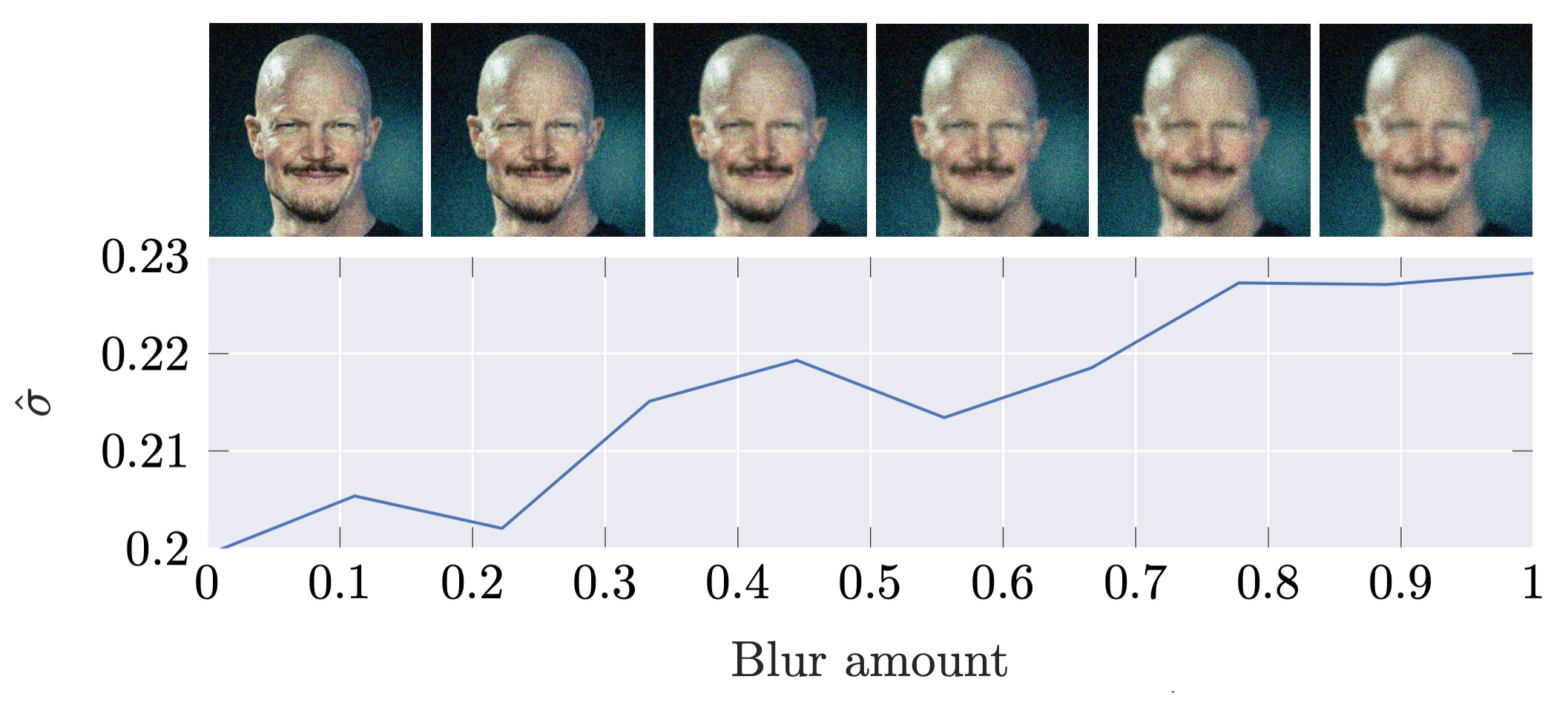}
		\caption{Measurement noise: $\sigma_{\vct{y}} = 0.1$}
		\label{fig:ordering_poor}
	\end{subfigure}
	\centering
	\begin{subfigure}[b]{0.7\textwidth}
		\includegraphics[width=1\linewidth]{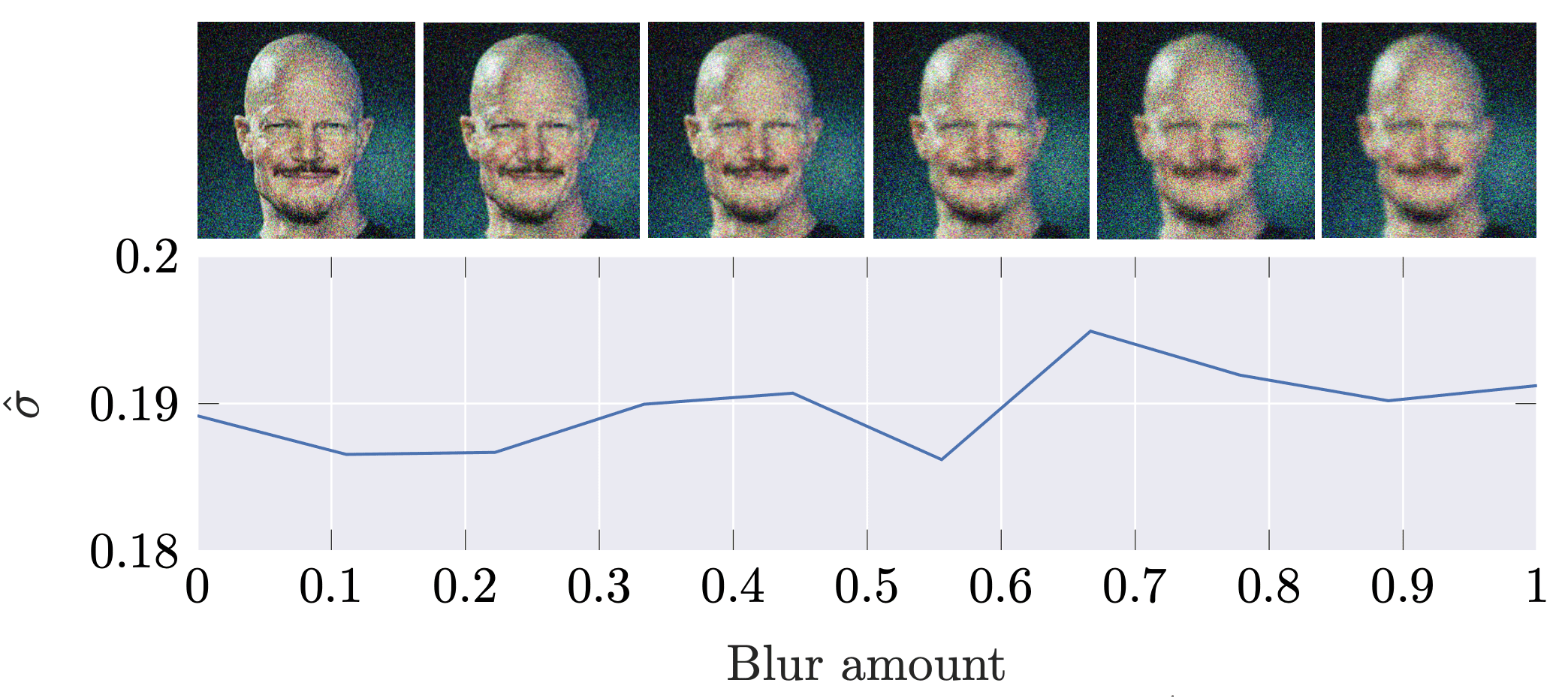}
		\caption{Measurement noise: $\sigma_{\vct{y}} = 0.2$}
		\label{fig:ordering_fail}
	\end{subfigure}
	\caption{Effect of perturbation in the measurement noise level in test time on the predictions of the severity encoder. We expect the predicted severity to be a non-decreasing function of the true blur amount, as depicted in Figure \ref{fig:ordering_correct}. At high noise level perturbations the severity encoder violates the monotonicity requirement (Figure \ref{fig:ordering_poor}) and eventually breaks down completely at extreme measurement noise variances (Figure \ref{fig:ordering_fail}).}\label{fig:ordering_complete}
\end{figure}

We demonstrate this property in Figure \ref{fig:ordering_plot} for the case where the measurement noise level corresponds to the training setting ($\sigma_{\vct{y}} = 0.05$). Here, we observe that the predicted severity is a non-decreasing function of the true blur level (monotonicity requirement is satisfied) when the ground truth clean image is fixed. Next, we increase the measurement noise in test time and investigate how the monotonicity requirement breaks down (see Figure \ref{fig:ordering_complete}).  At $\sigma_{\vct{y}} = 0.1$, which correspond to an increase in noise variance by a factor of $4\times$ compared to the training setting, the severity encoder fails to satisfy the monotonicity requirement (Fig. \ref{fig:ordering_poor}), but still manages to provide useful estimates in some regimes (high blur). This is due to the fact that (1) the severity encoder has not encountered such high noise levels during training and thus the estimates are highly inaccurate, and (2) the high noise 'over-powers' the effect of blurring and therefore it becomes more challenging to distinguish between images corrupted by similar blur amounts. Finally, as we increase the test-time measurement noise to $\sigma_{\vct{y}} = 0.2$ (variance $16\times$ training setup), the severity encoder completely breaks down, and produces a similar severity estimate for all blur amounts (Fig. \ref{fig:ordering_fail}).

To get an even more detailed understanding of the limitations of the severity encoder, we introduce the notion of \textit{ordering accuracy} that qualitatively captures how well the monotonicity requirement is satisfied. 
\begin{definition}[Ordering accuracy]
	Let $\vct{x}_0$ be a clean image, and $\vct{y}_{s_1}, \vct{y}_{s_2}, ..., \vct{y}_{s_n}$ a sequence of corrupted observations of $\vct{x}_0$, such that $\vct{y}_{s_i} =  \mathcal{A}(\vct{x}_0; s_i) + \vct{n}$, and $0 \leq s_1 < s_2 < ... < s_n \leq 1$, where $s_i$ parameterizes the severity level of the degradation (higher the more severe). Let $\hat{\sigma}_i := \hat{\sigma}(\vct{y}_{s_i})$ denote the corresponding estimates of degradation severity from the severity encoder, and 
\begin{equation}
	 g(\vct{y}_{s_i}, \vct{y}_{s_j}; \vct{x}_0)= \begin{cases} 1 & \text{if} ~~~ \text{sign}(s_i - s_j) = \text{sign}(\hat{\sigma}_i  - \hat{\sigma}_j) \\ 0 & \text{otherwise} \end{cases}
\end{equation}
	Then, the ordering accuracy (OA) of the severity encoder on sample $\vct{x}_0$ and severity levels $s_1,...,s_n$ under the degradation model $\mathcal{A}(\cdot, s)$ is 
	\begin{equation}
		\text{OA}(\vct{x}_0; \vct{y}_{s_1}, ..., \vct{y}_{s_n}) := \frac{1}{n (n-1)}\sum_{i \neq j} g(\vct{y}_{s_i}, \vct{y}_{s_j}; \vct{x}_0)
	\end{equation}
\end{definition}
Intuitively, the ordering accuracy takes a sequence of increasingly degraded images and the corresponding severity estimates from the encoder and evaluates the portion of pairs of degraded images for which the ordering of ground truth degradation level matches the ordering of predicted severities. Ideally, if the monotonicity requirement is satisfied, we have ordering accuracy of $1$ for every sample. On the other hand, if the severity estimates are generated randomly we have ordering accuracy of $0.5$ (the probability that any pair has the correct ordering is 0.5). 

\begin{figure}[htp]
	\centering
	\includegraphics[width=0.6\linewidth]{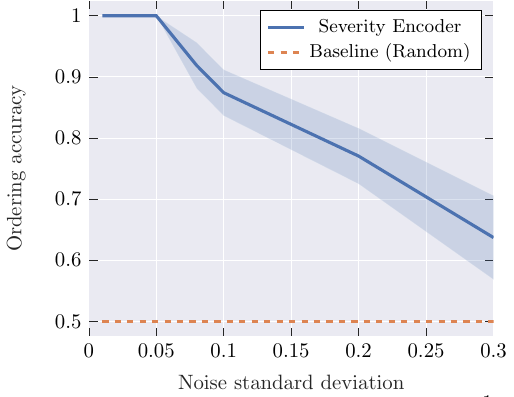}
	\caption{Evaluation of severity encoding performance under test-time noise level perturbation. The training noise level is $\sigma_{\vct{y}} = 0.05$. The severity encoder is robust to arbitrary reductions in noise level in test time (no drop in ordering accuracy). As the noise level perturbation increases, performance degrades and approaches random guessing (ordering accuracy of $0.5$). Mean over $100$ validation samples and $3$ measurement generating random seeds is plotted, with $\pm$ standard deviation over random seeds as shaded area.}\label{fig:ord_acc_vs_noise}
\end{figure}

We leverage the notion of ordering accuracy to evaluate the degradation of severity encoding performance under test-time measurement noise perturbations. In particular, we consider the Gaussian deblurring task and select $n=10$ blur kernel stds over a uniform grid in $[0.0, 3.0]$, with $s_1=0$ corresponding to no blurring and $s_n=1$ corresponding to blurring with blur kernel std of $3.0$. In Figure \ref{fig:ord_acc_vs_noise}, we plot how the average ordering accuracy evolves as we change the measurement noise level evaluated on $100$ validation samples on the CelebA dataset. We observe that for noise levels up to the training noise level, the severity encoder has perfect ordering accuracy. That is, we can expect reliable performance on noise levels that are different but lower than the training measurement noise. This property indicates that it may be beneficial to train the severity encoder at higher noise levels than the expected test-time noise in order to increase robustness.  As the noise level perturbation increases, severity encoding performance drops, approaching random guessing (ordering accuracy of $0.5$) at extreme noise variances.

\textbf{Robustness of \methodname{} performance --}Next, we investigate the impact of noise level mismatch in test-time on the reconstruction performance of adaptive solvers imbued with \methodname{}. In particular, we perform experiments on the FFHQ dataset with varying Gaussian blur and fixed measurement noise of $\sigma_{\vct{y}} = 0.05$ in severity encoder training. We pick ReSample as our baseline solver as it yields the best performance among non-adaptive solvers in our experiments. As the stopping criteria for hard data consistency ($\tau$ in \citet{song2023solving}) depends on the measurement noise, we expect degraded performance when the true noise level differs from the one used to tune the hyperparameters. We vary the measurement noise in test time and compare the reconstruction performance of \adaptive{ReSample} to ReSample and SwinIR. We depict the results of the robustness study in Figure \ref{fig:robustness}.  We observe that both in terms of perceptual and distortion metrics, the performance of \adaptive{ReSample} degrades more gracefully compared to its non-adaptive counterpart and SwinIR in most cases (with the exception of lower than expected noise and distortion metrics). We hypothesize that this additional robustness is due to the sample-by-sample adaptivity of \methodname{}. As the noise level increases, our method can automatically  adapt by increasing the number of diffusion steps due to higher degradation severity predictions from the severity encoder.

\begin{figure}[htp]
	\centering
	\includegraphics[width=0.59\linewidth]{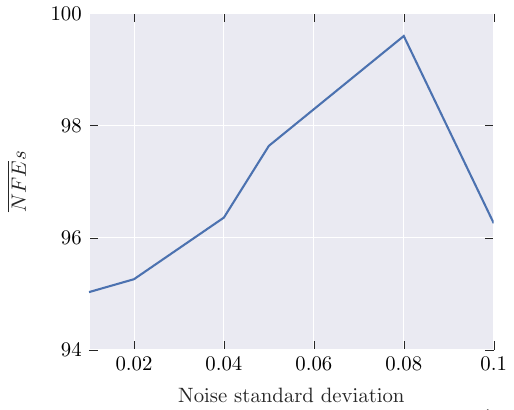}
	\caption{Due to the adaptivity of \methodname{}, the length of the sampling trajectory is automatically scaled to the  input noise level in test time, even in case of noise levels different from the training setup. At extreme noise discrepancies ($\sigma_{\vct{y}} = 0.1$) the severity encoder breaks down  and unable to adapt to the increased degradation severity.}\label{fig:robust_noise_steps}
\end{figure}

The adaptation to higher measurement noise levels in test time is demonstrated in Figure \ref{fig:robust_noise_steps}. Even though the severity encoder has not been trained on noise levels other than $\sigma_{\vct{y}} = 0.05$, its predictions are still useful enough to adapt the sampling trajectory to the changes in noise level, up to a limit. At  noise levels significantly higher than in the training setting ($\sigma_{\vct{y}} = 0.1$ and above), the performance of the severity encoder breaks down and is unable to provide accurate severity estimations. This phenomenon is corroborated by our observation of degrading ordering accuracy at high noise levels (see Figure \ref{fig:ord_acc_vs_noise} and Figure \ref{fig:ordering_poor}).

\color{black}
\subsection{Robustness against forward model mismatch}\label{apx:fwd_robustness}
Our method relies on a severity encoder that has been trained on paired data of clean and degraded images under a specific forward model.  We simulate a mismatch between the severity encoder fine-tuning operator and test-time operator in order to investigate the robustness of our technique with respect to forward model perturbations. 
In particular, we run the following experiments to assess the test-time shift: 1) we train the encoder on varying Gaussian blur and test on nonlinear blur and 2) we train the encoder on nonlinear blur and test on varying Gaussian blur. 

\begin{figure}[htp]
	\centering
	\includegraphics[width=0.7\linewidth]{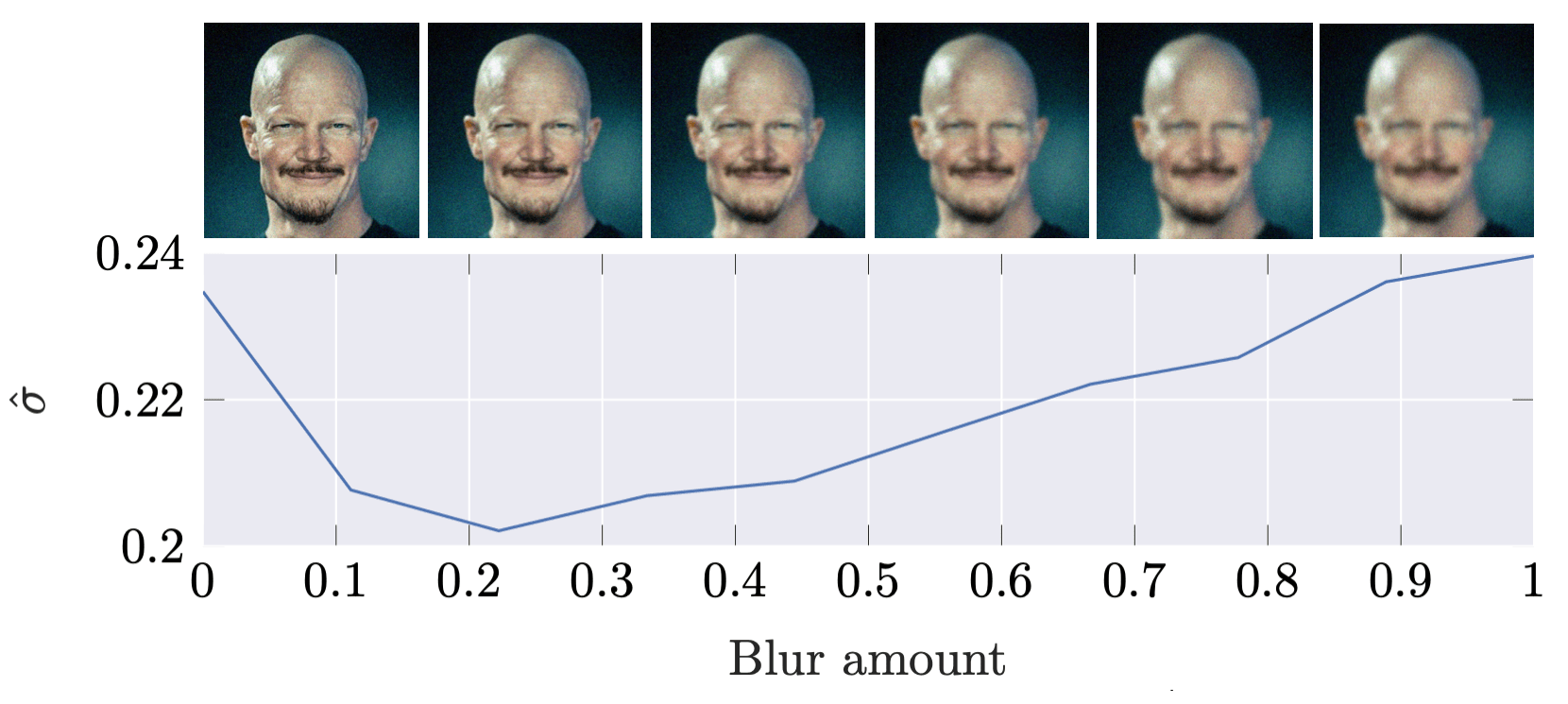}
	\caption{Effect of perturbation in the forward model in test time on the predictions of the severity encoder. We train the severity encoder on nonlinear blur and evaluate it on Gaussian blur of varying amounts. We expect the predicted severity to be a non-decreasing function of the true blur amount. However, due to the forward model mismatch the monotonicity requirement is violated for low blur levels. }\label{fig:ord_acc_vs_fwd}
\end{figure}

\textbf{Robustness of severity encoding --} First, following our methodology in Appendix \ref{apx:noise_robustness}, we investigate how severity encoding performance degrades if the image degradation forward model is different from what the severity encoder has been trained on. We train the severity encoder on nonlinear blur and evaluate it on Gaussian blur of varying amounts on the CelebA dataset. We expect the predicted severity to be a non-decreasing function of the true blur amount. However, as Figure \ref{fig:ord_acc_vs_fwd} demonstrates, the monotonicity requirement is violated due to the forward model mismatch. In particular, the severity encoder fails to predict the degradation severities of lightly blurred images. However, we observe satisfactory performance at high blur amounts (above $0.3$ in Fig. \ref{fig:ord_acc_vs_fwd}).  We hypothesize that at high blur levels, Gaussian blur and nonlinear blur become similar, thus the severity encoder trained on either may provide acceptable predictions for both. We further support this hypothesis by evaluating the ordering accuracy on a validation set of $100$ CelebA images. The results are summarized in Table \ref{tab:ord_acc_fwd}. We observe that the ordering accuracy is poor (mean $\text{OA} = 0.7704$) when evaluated across the full range of blur levels. However, it is near-perfect when low blur levels are excluded from the evaluation (mean $\text{OA} = 0.9778$). This experiment highlights the severity encoder's potential for generalization to unseen forward models, which opens up interesting future directions for fast adaptation to new tasks. 

\begin{table}[htp]
	\centering
	\begin{tabular}{|l|l|}
		\hline
		\textbf{Blur range}   & \textbf{Avg. OA }\\ \hline
		{[}0, 1{]}   & 0.7704  \\ \hline
		{[}0.3, 1{]} & 0.9778  \\ \hline
	\end{tabular}
	\caption{Ordering accuracy of a severity encoder trained on nonlinear blur and evaluated on Gaussian blur of varying blur amount. The severity encoder provides fair estimates even under forward model mismatch in the high blur regime.}\label{tab:ord_acc_fwd}
\end{table} 

\color{black}
\textbf{Robustness of \methodname{} performance --} We investigate the robustness of \adaptive{Latent-DPS} to forward model mismatch in test time. The results on the FFHQ test set are in Table \ref{tab:robust}. We observe minimal loss in performance when nonlinear blur encoder is used for reconstructing images corrupted by Gaussian blur. For the nonlinear deblurring task, using Gaussian blur encoder results in a more significant drop in the performance, while still providing acceptable reconstructions. These results are expected, as Gaussian blur can be thought of as a special case of the nonlinear blur model we consider. Therefore even when the encoder is swapped, it can provide meaningful mean and error estimation. However, the Gaussian blur encoder has never been trained on images corrupted by nonlinear blur. As such, the mean estimate is worse, resulting in a larger performance drop. Note that we did not re-tune the hyper-parameters in these experiments and doing so may potentially alleviate the loss in performance.
\begin{table}[htp]
	\centering
	\resizebox{16.5cm}{!}{
		\begin{tabular}[H]{ lccccccccc }
			\toprule
			&\multicolumn{4}{c}{\textbf{Gaussian Deblurring (varying)}} & &\multicolumn{4}{c}{\textbf{Nonlinear Deblurring}} \\
			\cmidrule{2-5}\cmidrule{7-10}
			\textbf{Method} &PSNR$(\uparrow)$&SSIM$(\uparrow)$&LPIPS$(\downarrow)$&FID$(\downarrow)$& &PSNR$(\uparrow)$&SSIM$(\uparrow)$&LPIPS$(\downarrow)$&FID$(\downarrow)$ \\
			\midrule
			{ \adaptive{Latent-DPS} + Gaussian blur encoder}&	{29.16}&{0.8191}&{0.2241}&{29.467}& &	{25.36}&	{0.7238}&	{0.3416}&	{54.90} \\
			\hline
			{ \adaptive{Latent-DPS} + NL blur encoder}&	{28.96}&	{0.8129}&	{0.2362}&	{30.34}& &	{27.22}&	{0.7705}&	{0.2694}&{36.92} \\
			\bottomrule 
		\end{tabular}
	}
	\caption{ Robustness experiments on the FFHQ test split. \label{tab:robust}}
\end{table}

\section{Experiments with Accelerated Baseline Solvers}\label{apx:ddim}

DDIM \citep{song2021denoising}, an accelerated sampling scheme, has demonstrated comparable generation performance to DDPM samplers with significantly lower compute requirements, often able to reduce the number of reverse diffusion steps from $N=1000$ to $N=100$ in the context of image synthesis. However, to the best of our knowledge, similar acceleration has not been successfully translated to latent domain diffusion solvers without significant drop in reconstruction performance. In this section, we investigate:
\begin{enumerate}
	\item \textbf{Non-adaptive accelerated solvers--} Can baseline solvers coupled with accelerated DDIM sampling match the reconstruction quality of \methodname{} adaptive counterparts with the same number of reverse diffusion steps? 
	\item \textbf{Adaptive accelerated solvers--} Can \methodname{} yield further speedup on top of accelerated solvers while maintaining reconstruction quality?
\end{enumerate}

 \textbf{Non-adaptive accelerated solvers-- }  
 We investigate whether the reduction in sampling steps of \methodname{} can be simply achieved using complete sampling trajectories with DDIM updates (without the shortcut initialization from the severity encoder). The DDIM updates in latent domain take the form
 \begin{equation}\label{eq:ddim_rev}
 	\vct{z}_{i-1} = \sqrt{\bar{\alpha}_{i-1}} \hat{\vct{z}}_0(\vct{z}_t) + \sqrt{1 - \bar{\alpha}_{i-1} - \delta^2 \tilde{\beta}_i^2}  \epsilon_{\vct{\theta}}(\vct{z}_i, i) +  \delta \tilde{\beta}_i^2 \vct{\epsilon},
 \end{equation}
 where $\hat{\vct{z}}_0(\vct{z}_t) $ is the estimator of the posterior mean based on Tweedie's formula, $ \tilde{\beta}_i = \sqrt{(1-\bar{\alpha}_{i-1}) / (1-\bar{\alpha}_{i})} \sqrt{1 - \bar{\alpha}_{i} / \bar{\alpha}_{i-1}}$ and $\delta$ parameter controlling the stochasticity of the sampling process. For $\delta = 0.0$ one obtains a fully deterministic inference procedure, while $\delta = 1.0$ corresponds to the ancestral sampling of DDPM. Here, we experiment with a variant of Latent-DPS, where we replace DDPM sampling ($N=1000$) with DDIM sampling of $N=100$, which we denote by Latent-DPS-DDIM($N=100$).  This closely matches the average number of steps \adaptive{Latent-DPS} uses on the varying Gaussian blur task on FFHQ, but instead of jumping ahead in the reverse process at initialization, it starts from the i.i.d. Gaussian distribution similar to DPS. We tune the data consistency step size $\eta$ in \eqref{apx:eta} by grid search over $[0.1, 0.5, 1.0, 1.5]$ on a validation split of $100$ images of FFHQ and select the value $\eta=1.5$. The results are shown in Table \ref{tab:ddim_ldps}. We find that Latent-DPS-DDIM($N=100$) far underperforms \adaptive{Latent-DPS} and other diffusion-based reconstruction techniques. In fact, we observe a large variation in reconstruction quality, from high fidelity reconstructions to images with heavy artifacts. We hypothesize that this effect is due to large errors in the posterior mean estimate in the early stages of diffusion resulting in inaccurate posterior sampling, which is further compounded by the large variance in sample degradations.
 
 Finally, we perform a similar experiment on ReSample, a more advanced solver that does not rely on DPS, but instead leverages hard data consistency. Table \ref{tab:ddim_resample} shows that ReSample performs significantly better than the naive Latent-DPS algorithm at high DDIM-acceleration ($N=50$), maintaining similar reconstruction quality to the non-accelerated original algorithm ($N=500$). However, the \methodname{} adaptive variant of ReSample far outperforms the DDIM-accelerated non-adaptive variant while using comparable number of reverse diffusion steps. These experiments further support that straightforward application of accelerated DDIM sampling in latent diffusion solvers, leveraging the same compute budget in inference time, cannot compete with \methodname{} adaptive variants in terms of reconstruction quality.

 \textbf{Adaptive accelerated solvers-- } Our proposed framework can be paired with any diffusion-based sampling technique as long as the SNR can be evaluated analytically for any time step (see \eqref{eq:snr_match}). Here, we investigate the performance of \methodname{}, when paired with highly accelerated DDIM sampling. We run experiments on the FFHQ test set under Gaussian blur degradation of varying magnitude matching the setup in Section \ref{sec:exp}. We investigate the performance of adaptive versions of Latent-DPS-DDIM($N=20$) and Latent-DPS-DDIM($N=100$) for both $\delta=0.0$ (deterministic) and $\delta=1.0$ (DDPM-like).  We tune the data consistency step size $\eta$ in \eqref{apx:eta} by grid search over $[0.1, 0.5, 1.0, 1.5]$ on a validation split of $100$ images of FFHQ for each combination of $N$ and $\delta$. We select the value $\eta=1.5$ as it provides the best LPIPS across all experiments.

The results are shown in Table \ref{tab:ddim_ours}.  First, we observe that combining \methodname{} with DDIM sampling results in vastly accelerated sampling: we obtain reasonable reconstructions in $10.4$ ($N=100$) or $2.5$ ($N=20$) diffusion steps on average over the test set. Image quality in terms of distortion and perceptual metrics is comparable with other diffusion based solvers under DDIM sampling with N=100 steps, but with compute cost reduced by an order of magnitude. However, reconstruction quality degrades significantly when deploying Latent-DPS-DDIM($N=20$). We hypothesize that the LDPS step size scheduling defined in \eqref{apx:eta} needs to be tailored to DDIM sampling in this regime, which is an interesting direction for future work. Finally, we note that we find \methodname{} performance to be robust to the setting of $\delta$ in the DDIM sampler, as we obtain very similar results for $\delta = 0.0$ and $\delta=1.0$.

\begin{table}[htp]
	\centering
	\resizebox{16.5cm}{!}{
		\begin{tabular}[H]{ lccccccccc }
			\toprule
			&\multicolumn{4}{c}{\textbf{Gaussian Deblurring (varying)}} & &\multicolumn{4}{c}{\textbf{Nonlinear Deblurring}} \\
			\cmidrule{2-5}\cmidrule{7-10}
			\textbf{Method} &PSNR$(\uparrow)$&SSIM$(\uparrow)$&LPIPS$(\downarrow)$&FID$(\downarrow)$& &PSNR$(\uparrow)$&SSIM$(\uparrow)$&LPIPS$(\downarrow)$&FID$(\downarrow)$ \\
			\midrule
			{Latent-DPS-DDIM($N=100$), $\delta=0.0$}&	{16.62}&{0.4192}&{0.6224}&{233.74}& &	{13.52}&	{0.3694}&	{0.6472}&	{278.25} \\
			{Latent-DPS-DDIM($N=100$), $\delta=1.0$}&	{18.23}&	{0.4771}&	{0.5474}&	{129.67}& &	{14.96}&	{0.4071}&	{0.5533}&{86.68} \\
			\bottomrule 
		\end{tabular}
	}
	\caption{ Latent-DPS without severity encoding using DDIM sampling ($N=100$) on the FFHQ test split.  \label{tab:ddim_ldps}}
\end{table}

\begin{table}[htp]
	\Huge
	\centering
	\renewcommand{\arraystretch}{1.3} 
	\resizebox{16.5cm}{!}{
		\begin{tabular}[H]{ lccccccccccccccccc }
			\toprule
			&\multicolumn{5}{c}{\textbf{Gaussian Deblurring (varying)}} & &\multicolumn{5}{c}{\textbf{Nonlinear Deblurring}} & &\multicolumn{5}{c}{\textbf{Random Inpainting}}\\
			\cmidrule{2-6}\cmidrule{8-12}\cmidrule{14-18}
			\textbf{Method} &PSNR$(\uparrow)$&SSIM$(\uparrow)$&LPIPS$(\downarrow)$&FID$(\downarrow)$& $\overline{\text{NFE}}$& &PSNR$(\uparrow)$&SSIM$(\uparrow)$&LPIPS$(\downarrow)$&FID$(\downarrow)$ &$\overline{\text{NFE}}$&&PSNR$(\uparrow)$&SSIM$(\uparrow)$&LPIPS$(\downarrow)$&FID$(\downarrow)$ &$\overline{\text{NFE}}$ \\
			\midrule
			{ReSample(DDIM, N=500)}&	28.77&0.8219&0.2587&81.96&500&  &26.62&	0.7318&	0.2838&	68.57& 500&&27.51&	0.7892&	0.2460&	63.39& 500  \\
			ReSample(DDIM, N=50)&28.15& 0.7910& 0.2641& 83.05& 50& & 25.59& 0.6810& 0.3697& 118.40& 50 & & 26.95& 0.7525& 0.3154& 115.42& 50  \\
			\hline
			\adaptive{ReSample(DDIM, N=500)}&\textbf{29.07}& \textbf{0.8330}& \textbf{0.2383}& \textbf{74.76}& 49.90& & \textbf{26.88}& \textbf{0.7660}& \textbf{0.2667}& \textbf{64.57}& 67.81& & \textbf{28.13}& \textbf{0.8260}& \textbf{0.2045}& \textbf{56.67}& 52.09  \\
			\bottomrule 
		\end{tabular}
	}
	\caption{Comparison of ReSample and its DDIM-accelerated variant with \methodname{} adaptation on the FFHQ test split.  \label{tab:ddim_resample}}
\end{table}

\begin{table}[htp]
	\centering
	\resizebox{16.5cm}{!}{
		\begin{tabular}[H]{ lcccccccccc }
			\toprule
			& &\multicolumn{4}{c}{\textbf{$\delta=0.0$}} & &\multicolumn{4}{c}{\textbf{$\delta=1.0$}} \\
			\cmidrule{3-6}\cmidrule{8-11}
			\textbf{Method}& Avg. NFE &PSNR$(\uparrow)$&SSIM$(\uparrow)$&LPIPS$(\downarrow)$&FID$(\downarrow)$& &PSNR$(\uparrow)$&SSIM$(\uparrow)$&LPIPS$(\downarrow)$&FID$(\downarrow)$ \\
			\midrule
			{\adaptive{Latent-DPS}}& 100.1&	-&-&-&-& &	{29.16 }&	{0.8191}&	{0.2241}&	{29.46} \\
			{\adaptive{Latent-DPS-DDIM($N=100$)}}& 10.4&	{28.04}&	{0.7956}&	{0.2537}&	{32.08}& &	{27.70}&	{0.7901}&	{0.2550}&{31.83} \\
			{\adaptive{Latent-DPS-DDIM($N=20$)}}& 2.5&	{26.05}&	{0.7482}&	{0.3204}&	{57.77}& &	{26.09}&	{0.7504}&	{0.3168}&{54.88} \\
			\bottomrule 
		\end{tabular}
	}
	\caption{Comparison of DDIM-accelerated sampling techniques in the \methodname{} framework on FFHQ varying Gaussian deblurring. \label{tab:ddim_ours}}
\end{table}

\section{Adaptive DDIM Encoding Initialization}\label{apx:init}
So far we have initialized the sampling trajectory with a noise-corrected latent estimate obtained from our severity encoder, where the noise correction is aimed at suppressing structure in the estimation error. This initialization scheme is analogous to stochastic encoding \citep{song2023solving}. In this section, we investigate further techniques to initialize the reverse diffusion process in our framework in a sample-adaptive fashion.

With DDIM, the generative process can be reversed deterministically by simply running 
\begin{equation}\label{eq:ddim_fwd}
	\vct{z}_{i+1} = \sqrt{\alpha_i} \vct{z}_{i} - \sqrt{1 - \alpha_i} \sqrt{1 - \bar{\alpha}_i} s_{\vct{\theta}}(\vct{z}_i, i), 
\end{equation}
which can be obtained from reversing \eqref{eq:ddim_rev}. Here, $s_{\vct{\theta}}$ denotes the learned score (scaled version of $\epsilon_{\vct{\theta}}$).  When starting from the clean latent $\vct{z}_0$ and running \eqref{eq:ddim_fwd} for $t = 0,..,N-1$ one obtains a deterministic encoding of $\vct{z}_0$ and the process is referred to as DDIM encoding \citep{preechakul2022diffusion}. STSL \citep{rout2023beyond} leverages DDIM encoding of a naive reconstruction ($\enc{\mtx{A}^T \vct{y}}$, where $\varepsilon_0$ denotes the original encoder pretrained with the LDM) to obtain an initialization, where the DDIM forward process is run from $t=0$ to $T$. This initialization scheme can be combined with Flash adaptation, that is we can perform “partial” DDIM encoding from $t=0$ to $t_{start}$, where $t_{start}$ is the starting time index predicted by severity encoding. We refer to this technique as \textit{adaptive DDIM encoding}. We perform experiments for the following scenarios:
\begin{itemize}
\item \textbf{STSL initialization with adaptation:} we use adaptive DDIM encoding of $\enc{\mtx{A}^T\vct{y}}$ by running the DDIM forward process for $t_{start}$ steps, where $t_{start}$ is obtained from the severity encoder. We do not use noise correction.

\item \textbf{Flash initialization with DDIM encoding:} we use adaptive DDIM encoding of $\hat{z}$ (latent estimate from severity encoder) by running the DDIM forward process for $t_{start}$ steps, where $t_{start}$ is obtained from the severity encoder. We do not use noise correction.
\end{itemize}
We choose Latent-DPS with DDIM sampling ($N=500$, $\eta=0.0$) as our baseline solver (same setting as described in Appendix \ref{apx:ddim}) in order to make the reverse process as close to DDIM sampling as possible within our framework. We show results on the FFHQ test set with varying Gaussian blur and random inpainting operators in Table \ref{tab:ddim_enc}. The mean $t_{start}$ predicted by severity encoding on the dataset is $49.91$ for deblurring and $52.08$ for inpainting. 

\begin{table}[h!]
	\centering
		\resizebox{16.5cm}{!}{
	\begin{tabular}{lccccccccc}
		\toprule
		&\multicolumn{4}{c}{\textbf{Gaussian Deblurring (varying)}} & &\multicolumn{4}{c}{\textbf{Random Inpainting}} \\	\cmidrule{2-5}\cmidrule{7-10}
		\textbf{Initialization} & SSIM$(\uparrow)$  & PSNR$(\uparrow)$  & LPIPS$(\downarrow)$ & FID$(\downarrow)$& &  SSIM$(\uparrow)$  & PSNR$(\uparrow)$  & LPIPS$(\downarrow)$ & FID$(\downarrow)$   \\ \midrule
		$\hat{\vct{z}}$ + noise correction (ours) & 0.8136 & 28.96 & 0.2328 & 56.91 & &0.8447 & 29.11 & 0.2010 & 57.24 \\ 
		$\hat{\vct{z}}$ + DDIM encoding           & 0.8488 & 30.05 & 0.2497 & 82.54 & &0.8605 & 29.38 & 0.2132 & 68.23   \\ 
		$\enc{\mtx{A}^T \vct{y}}$ + DDIM encoding     & 0.7906 & 28.12 & 0.3201 & 92.08 & &0.6337 & 24.87 & 0.4302 & 135.4 \\ \bottomrule
	\end{tabular}
}
	\caption{Results on the FFHQ dataset using various initialization schemes.}
	\label{tab:ddim_enc}
\end{table}


First, we observe that STSL initialization with adaptation performs worse than our vanilla Flash initialization. This is natural, as we leverage a supervised model to obtain an initial reconstruction that is of better quality than the unsupervised $\enc{\mtx{A}^T\vct{y}}$ initialization. Furthermore, we suspect that since the autoencoder of LDM has been pretrained predominantly on clean images, it may compress $\mtx{A}^T \vct{y}$ inefficiently.

Second, we find that adaptive DDIM encoding on top of the latent estimate $\hat{\vct{z}}$ obtained from our severity encoder performs surprisingly well. Even though this initialization falls short in terms of perceptual quality compared to vanilla Flash initialization with noise correction, it achieves superior distortion (PSNR, SSIM). This hints at blurrier, but highly reliable reconstructions faithful to the original measurement. This experiment suggests that \methodname{} combined with adaptive DDIM encoding initialization may be very promising. In particular, by tuning a hyperparameter that controls the number of DDIM forward steps for encoding (similar to our noise correction parameter $c$) these results may be further improved. We leave this exploration for future work.

We note that a negative side-effect of using adaptive DDIM encoding in our setting is that it may limit the baseline solver to DDIM variants, as it is unclear how this initialization scheme performs when combined with arbitrary sampling techniques. 

\section{Additional Efficiency Experiments}
\begin{figure}[htp]
	\centering
		\begin{subfigure}{0.33\textwidth}
				\centering
				\includegraphics[width=0.88\textwidth]{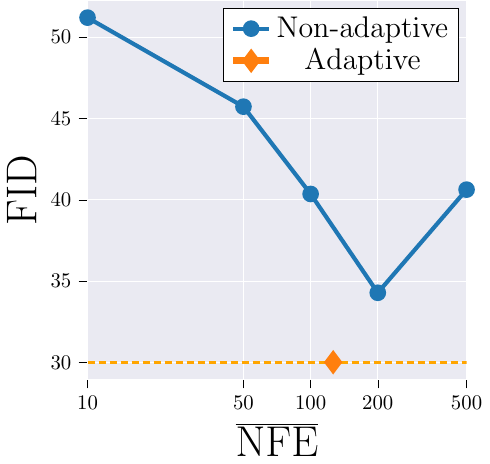}
			\end{subfigure}%
		\begin{subfigure}{0.33\textwidth}
				\centering
				\includegraphics[width=0.9\textwidth]{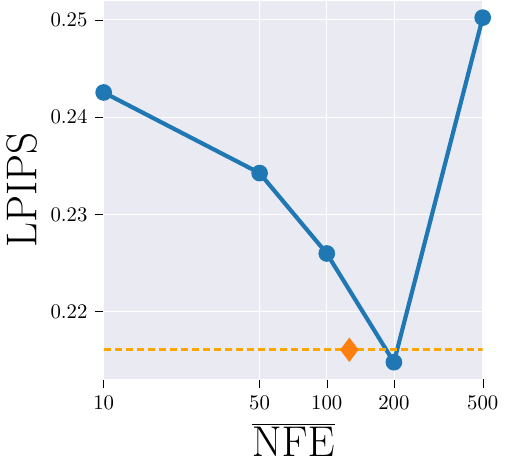}
			\end{subfigure}%
		\begin{subfigure}{0.33\textwidth}
				\centering
				\includegraphics[width=0.88\textwidth]{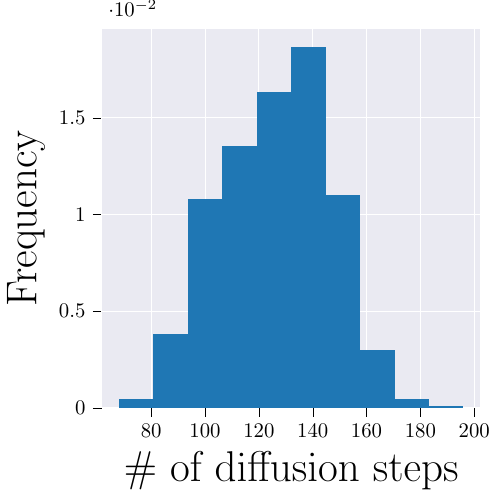}
			\end{subfigure}
	\caption[short]{Comparison of \adaptive{Latent-DPS} with the non-adaptive counterpart with fixed starting time for all samples (nonlinear deblurring, CelebA-HQ). \uline{Left and center}: We plot reconstruction quality as a function of the number of diffusion steps for the non-adaptive variant with various fixed starting times. The data point corresponding to our adaptive method depicts the average number of diffusion steps due to adaptive starting time.  \methodname{} achieves the best FID and near-optimal LPIPS compared to any choice of non-adaptive starting time. \uline{Right}: Histogram of predicted starting times for \adaptive{Latent-DPS}.\vspace{-0.5cm}\label{fig:nl_nfe_efficiency}}
\end{figure}

We perform additional experiments on the nonlinear blurring task to demonstrate the efficiency of \methodname{} compared to non-adaptive baselines. Figure \ref{fig:nl_nfe_efficiency} indicates that the adaptive sampling of \methodname{} achieves better perceptual quality than any fixed-length sampling trajectory. The adaptivity is further supported by the wide spread of sampling steps utilized by \methodname{}.

\color{black}
\section{Limitations}\label{apx:limitations}
We identify the following limitations of our framework. 
\begin{enumerate}
	\item The proposed reconstruction method requires degraded-clean image pairs for fine-tuning the severity encoder. Fine-tuning has to be performed separately for each degradation, thus the method is less flexible than DPS and similar diffusion solvers. However, we argue that the fine-tuning step has fairly low cost and greatly pays off in reconstruction performance and efficiency. Moreover, investigating the viability of a general degradation severity estimator that works for arbitrary degradations (under some reasonable assumptions) is an interesting direction for future research.
	\item The proposed severity estimation method breaks down at  high noise perturbations compared to the training settings and when there is a significant test-time shift in the forward model. 
	\item The assumption of i.i.d. Gaussian prediction error provides a simple way to estimate the severity, however does not necessarily hold in practice. We believe that more realistic error models can further improve our technique, which we leave for future work.
\end{enumerate}  

\newpage
\section{Additional Samples for Visual Comparison}\label{sec:vis_comp}
\begin{figure}[H]
	\centering
	\includegraphics[width=0.95\linewidth]{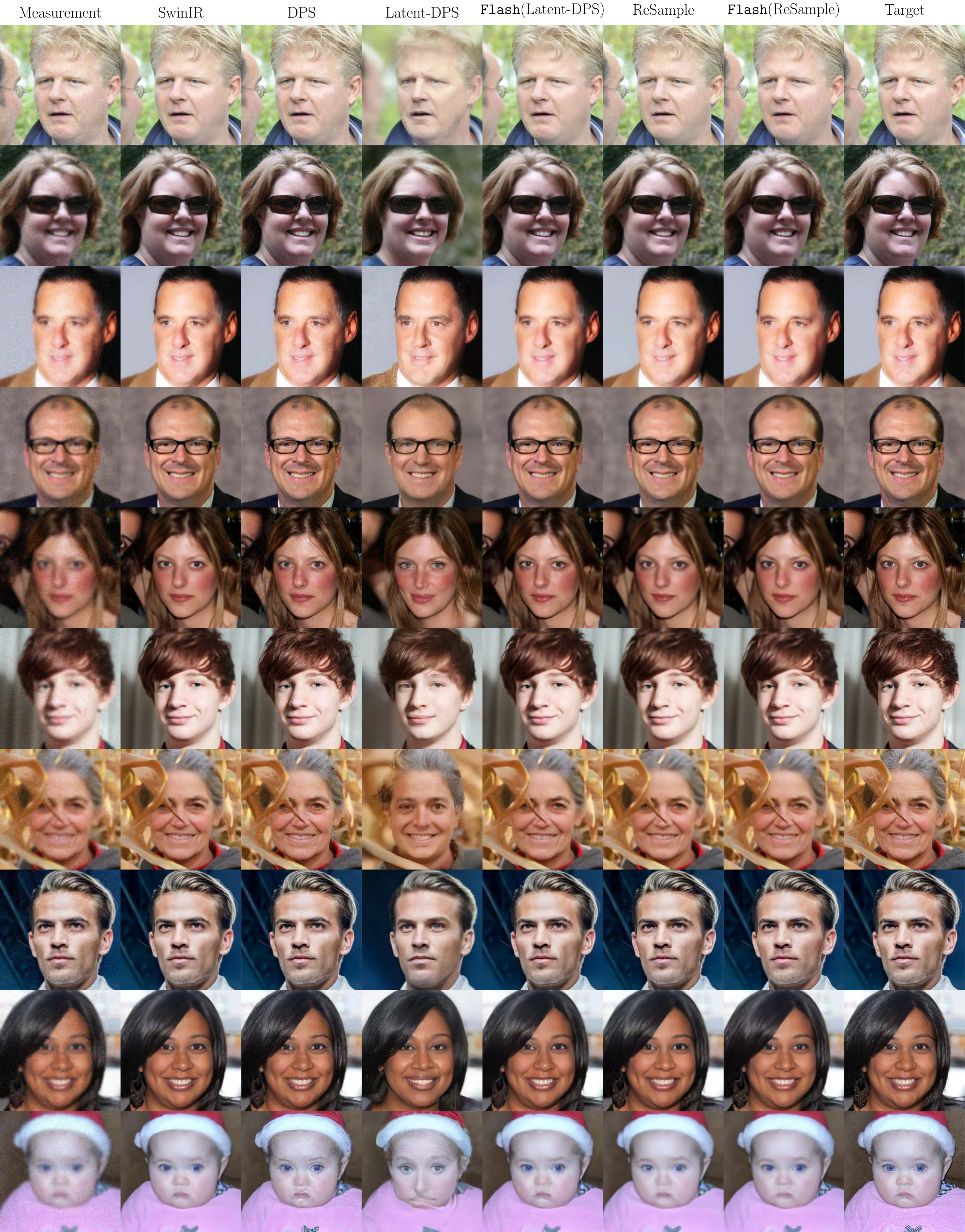}
	\caption{Reconstructed samples from the FFHQ test set on varying Gaussian blur with additive noise $\sigma_{\vct{y}} = 0.05$. The samples are not cherry-picked.}
\end{figure}
\begin{figure}[htp]
	\centering
	\includegraphics[width=0.95\linewidth]{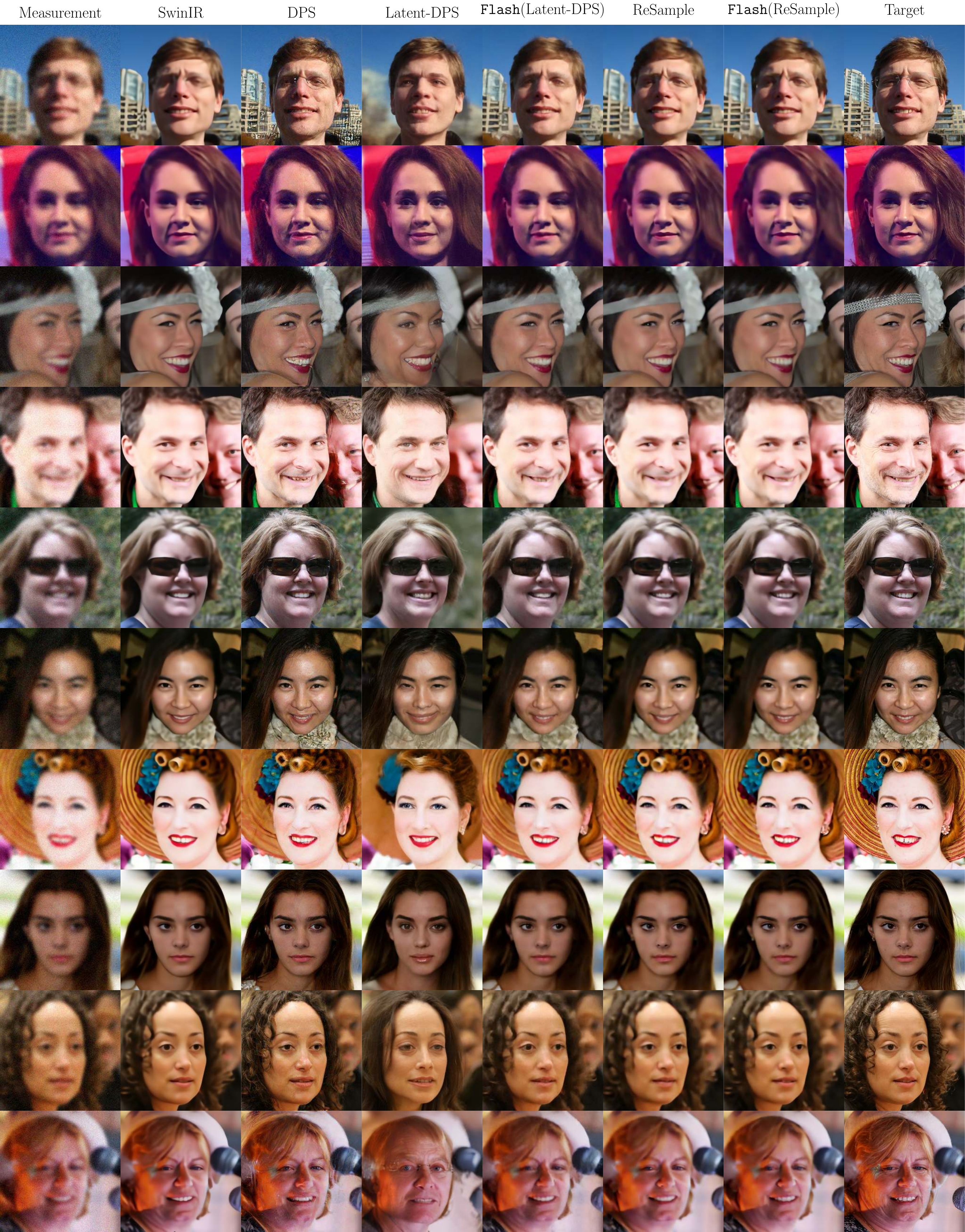}
	\caption{Reconstructed samples from the FFHQ test set on fixed Gaussian blur with additive noise $\sigma_{\vct{y}} = 0.05$. The samples are not cherry-picked.}
\end{figure}
\begin{figure}[htp]
	\centering
	\includegraphics[width=0.95\linewidth]{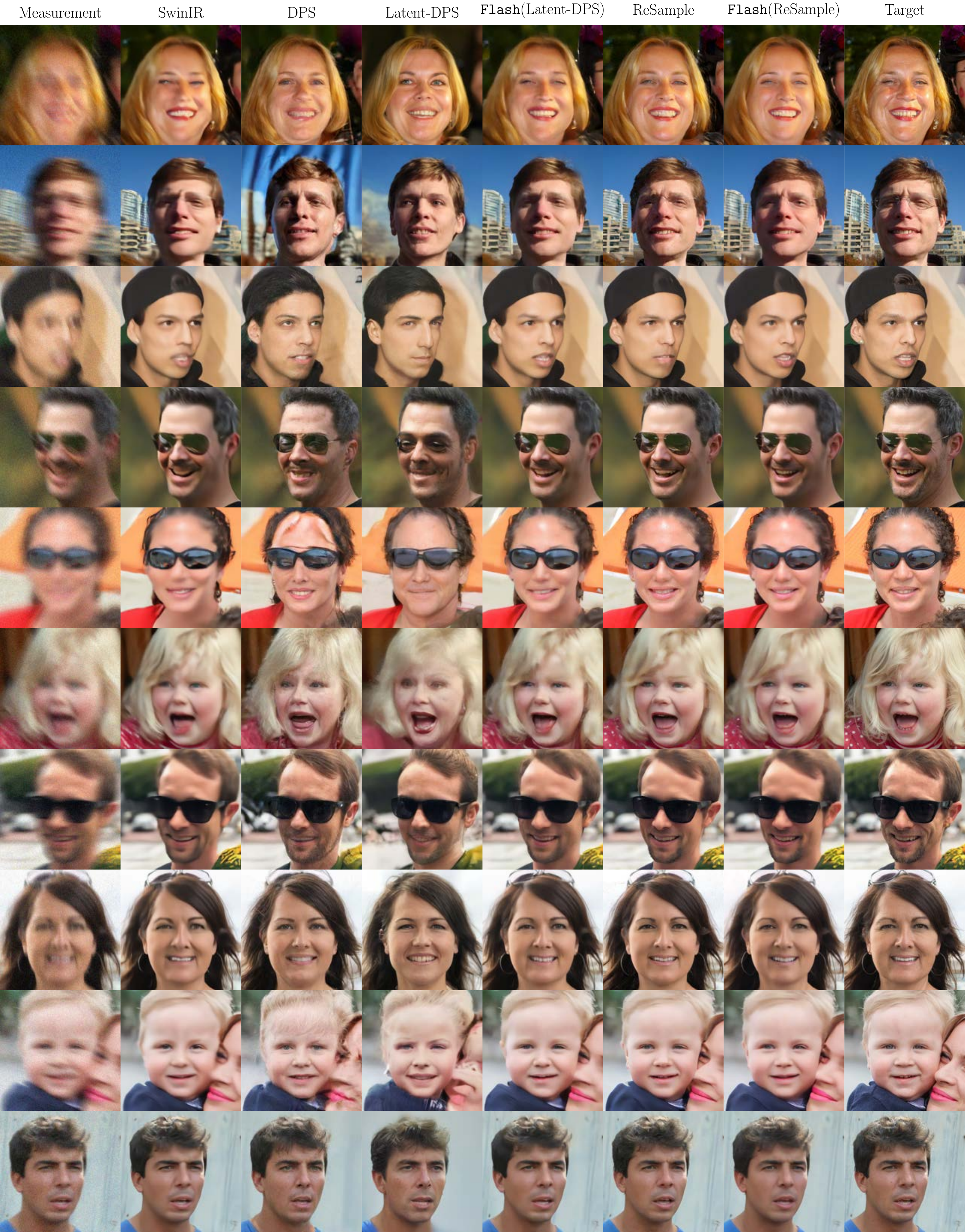}
	\caption{Reconstructed samples from the FFHQ test set on nonlinear motion blur with random kernels and additive noise $\sigma_{\vct{y}} = 0.05$. The samples are not cherry-picked.}
\end{figure}
\begin{figure}[htp]
	\centering
	\includegraphics[width=0.95\linewidth]{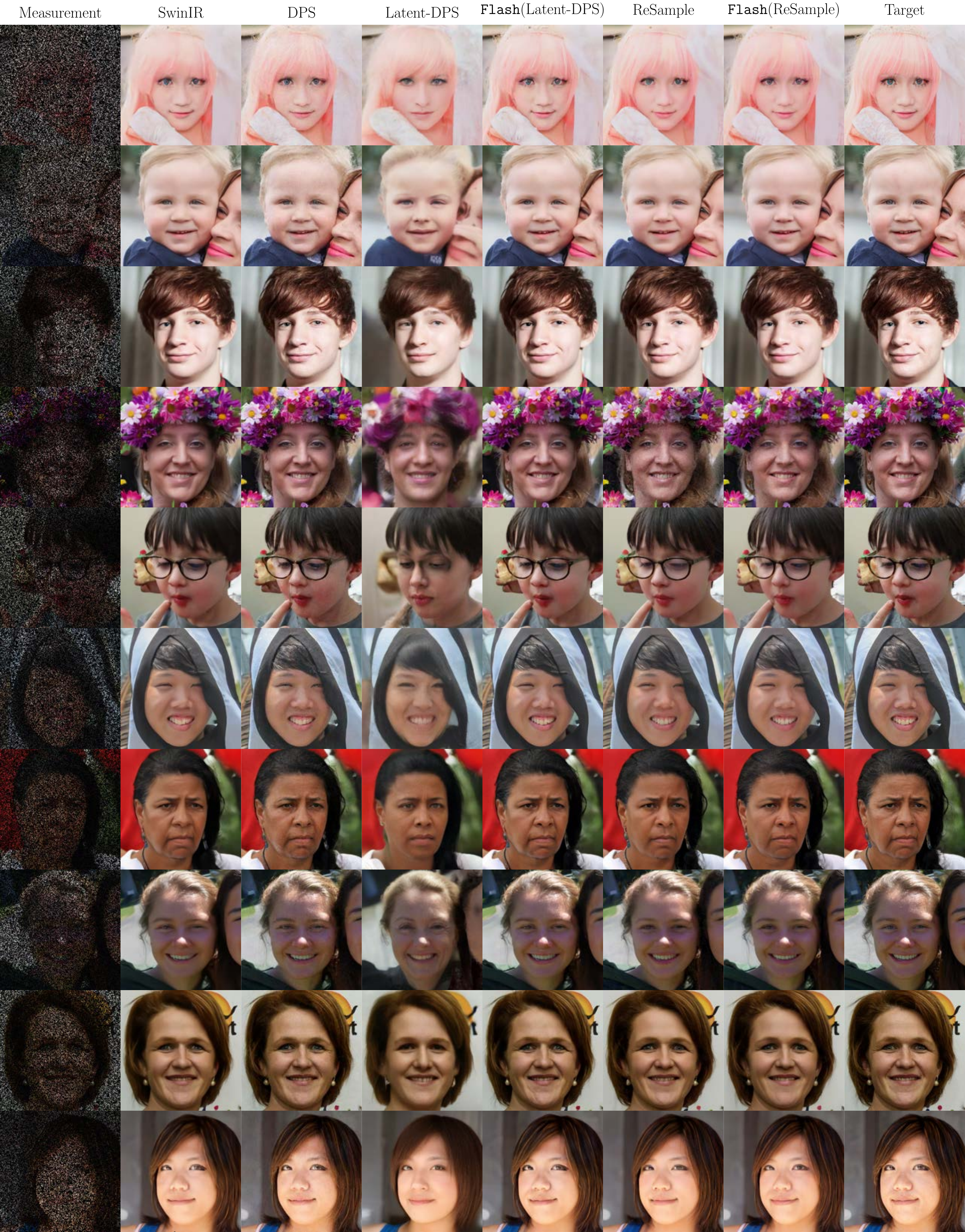}
	\caption{Reconstructed samples from the FFHQ test set on varying amounts of random inpainting with additive noise $\sigma_{\vct{y}} = 0.05$. The samples are not cherry-picked.}
\end{figure}
\begin{figure}[htp]
	\centering
	\includegraphics[width=0.95\linewidth]{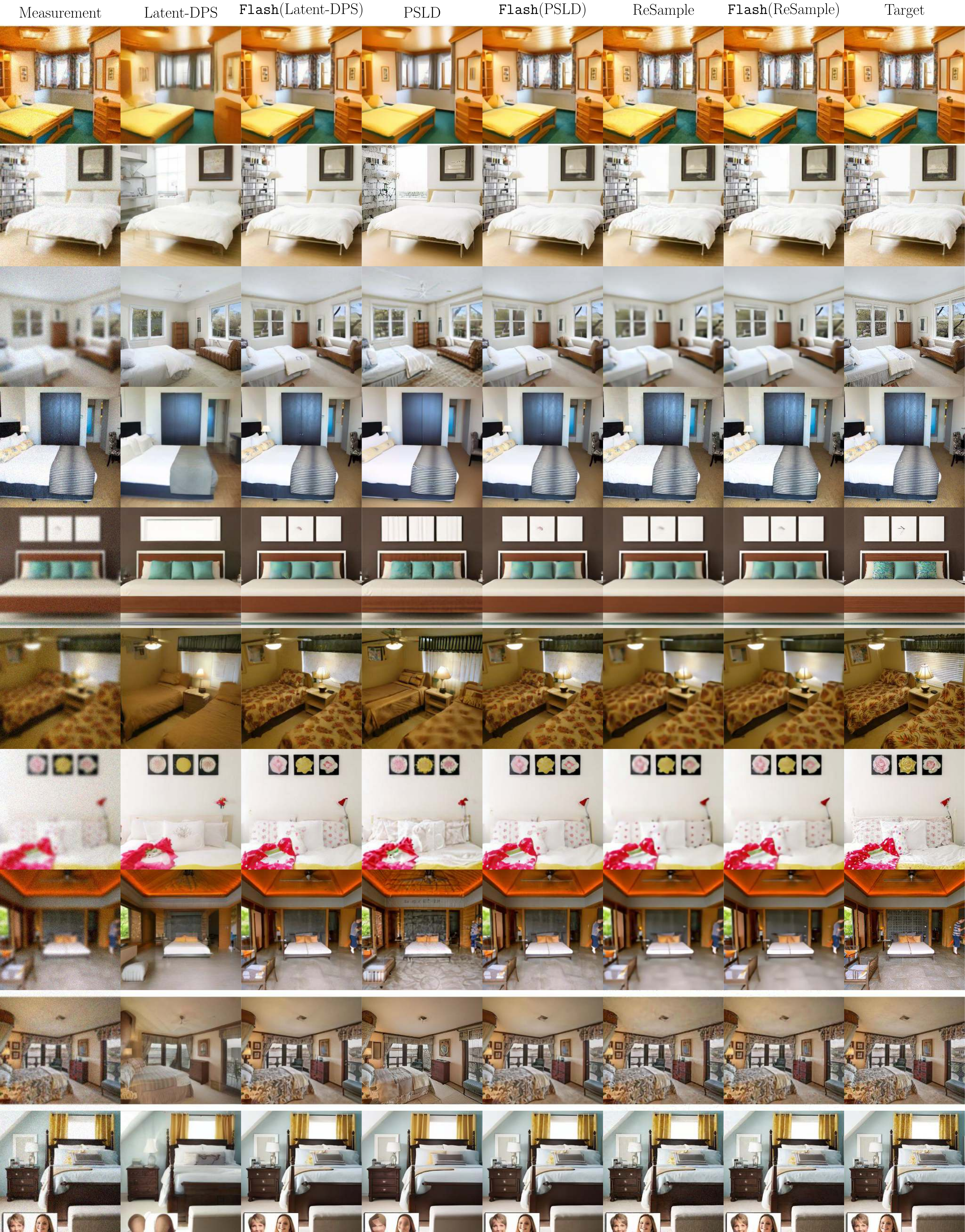}
	\caption{Reconstructed samples from the LSUN Bedrooms test set on varying Gaussian blur with additive noise $\sigma_{\vct{y}} = 0.05$. The samples are not cherry-picked.}
\end{figure}
\begin{figure}[htp]
	\centering
	\includegraphics[width=0.95\linewidth]{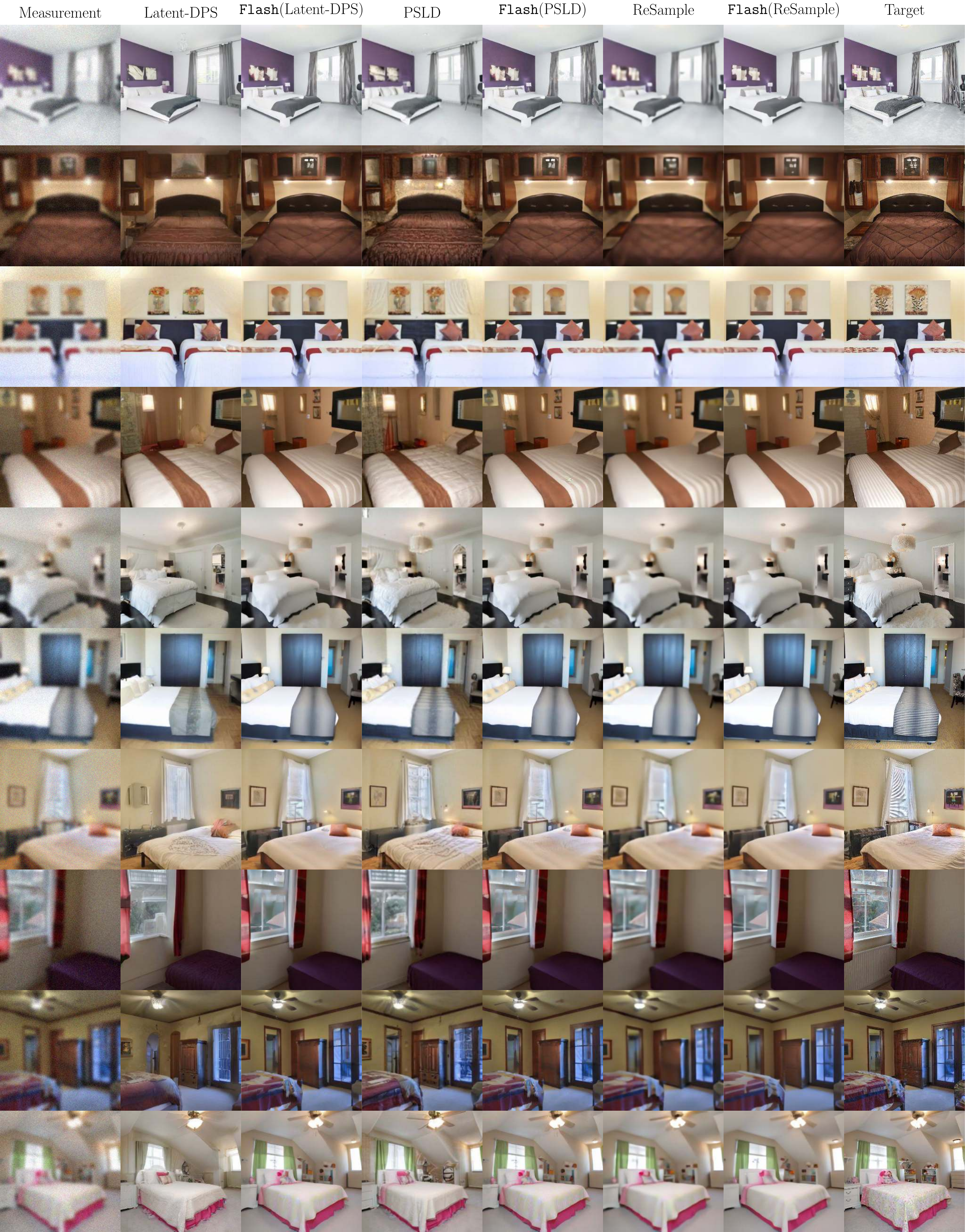}
	\caption{Reconstructed samples from the LSUN Bedrooms test set on fixed Gaussian blur with additive noise $\sigma_{\vct{y}} = 0.05$. The samples are not cherry-picked.}
\end{figure}
\begin{figure}[htp]
	\centering
	\includegraphics[width=0.95\linewidth]{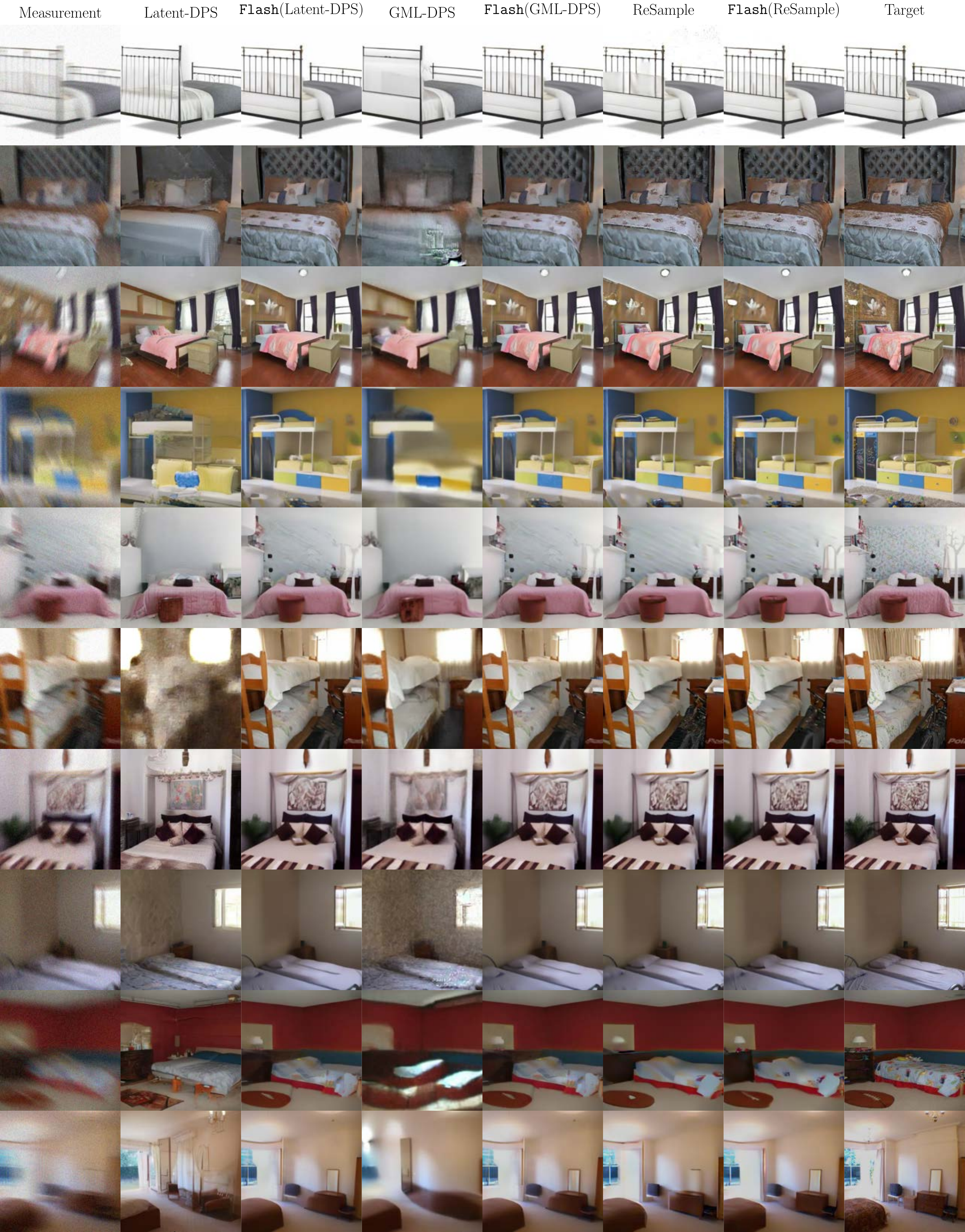}
	\caption{Reconstructed samples from the LSUN Bedrooms test set on nonlinear motion blur with random kernels and additive noise $\sigma_{\vct{y}} = 0.05$. The samples are not cherry-picked.}
\end{figure}
\begin{figure}[htp]
	\centering
	\includegraphics[width=0.95\linewidth]{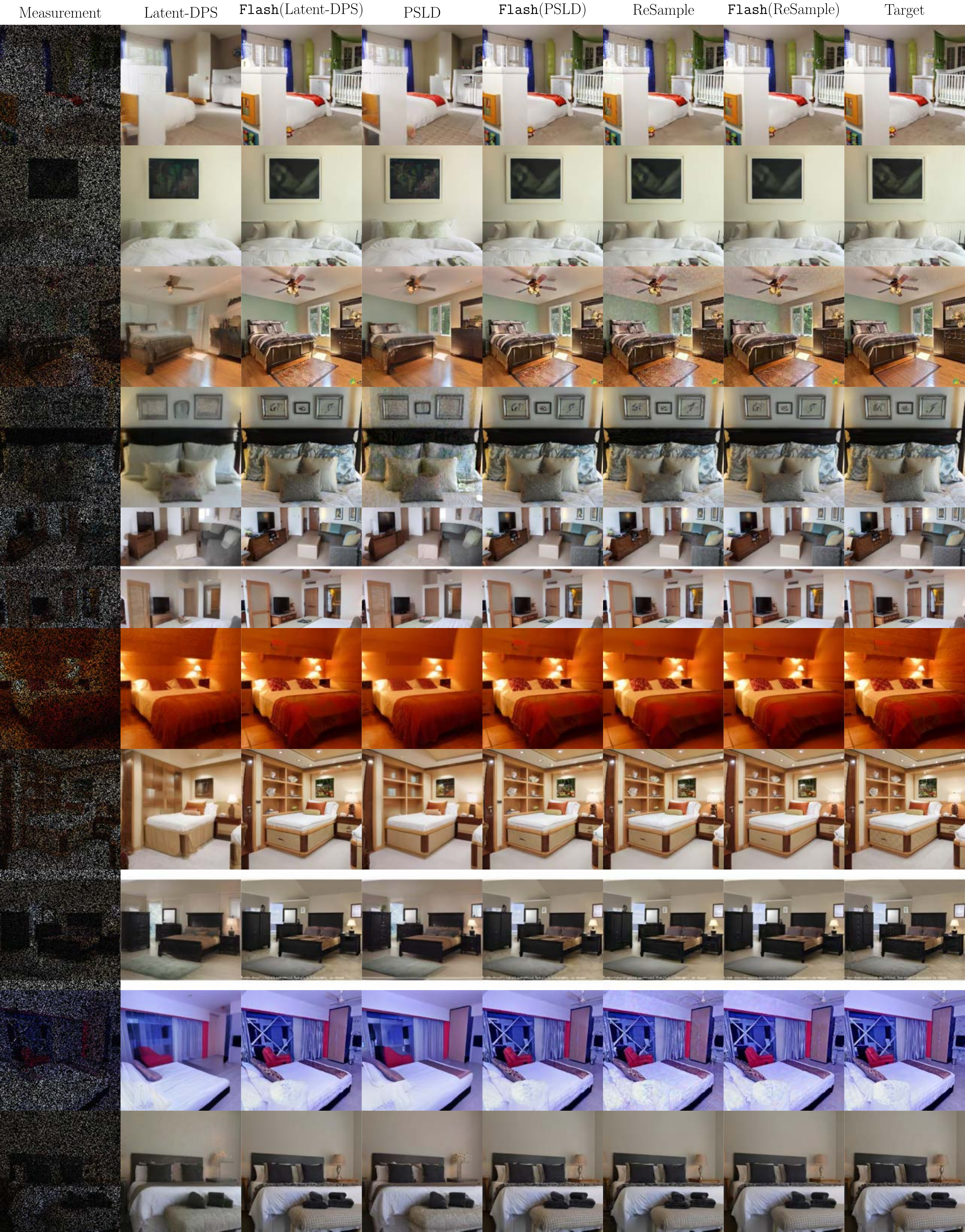}
	\caption{Reconstructed samples from the LSUN Bedrooms test set on varying amounts of random inpainting with additive noise $\sigma_{\vct{y}} = 0.05$. The samples are not cherry-picked.}
\end{figure}

\end{document}